# Unidirectional magnetoelectric-field multiresonant tunneling

E.O. Kamenetskii, E. Hollander, R. Joffe, and R. Shavit

Microwave Magnetic Laboratory,
Department of Electrical and Computer Engineering,
Ben Gurion University of the Negev, Beer Sheva, Israel

July 17, 2014

**Abstract**

Unidirectional multiresonant tunneling of the magnetoelectric (ME)-field excitations through a subwavelength (regarding the scales of regular electromagnetic radiation) vacuum or isotropic-dielectric regions has been observed in two-port microwave structures having a quasi-2D ferrite disk with magnetic-dipolar-mode (MDM) oscillations. The excitations manifest themselves as the Fano-resonance peaks in the scattering-matrix parameters at the stationary states of the MDM spectrum. The ME near-field excitations are quasimagnetostatic fields with non-zero helicity parameter. Topological-phase properties of ME fields are determined by edge chiral currents of MDM oscillations. We show that while for a given direction of a bias magnetic field (in other words, for a given direction of time), the ME-field excitations are considered as "forward" tunneling processes, in the opposite direction of a bias magnetic field (the opposite direction of time), there are "backward" tunneling processes. Unidirectional ME-field resonant tunneling is observed due to distinguishable topology of the "forward" and "backward" ME-field excitations. We establish a close connection between the Fano-resonance unidirectional tunneling and topology of ME fields in different microwave structures.

## I. INTRODUCTION

Topological phases have been attracting much attention in various fields in physics. In condensed matters, this leads to the foundation of topological isolators. As one of the interesting examples of such structures there are photonic crystals with chiral edge states. It was proposed that due to these edge states there should be underactional propagation of electromagnetic energy [1 – 3]. Recently, unidirectional (chiral) edge modes of magnetostatic (MS) [or magnetic-dipolar-mode (MDM)] waves were found in a magnonic crystal [4, 5]. It was discussed that in such structures, the magnetic dipolar interaction joins the relative rotational angle between the spin space and orbital space. It was shown that the chiral edge MS modes can break both the time-reversal symmetry and reflection symmetry and can propagate in a direction opposite to the Damon-Eschbach surface MS modes [6].

Recent studies reveal that unidirectional (chiral) edge states of magnetization can also be exhibited in another type of a ferrite structure – a single ferrite-disk particle with a MDM spectrum [7 – 13]. At the MDM resonances in a quasi-2D ferrite disk, together with spinning rotation of elementary magnetic dipoles also orbital rotation of the entire-structure topological magnetic dipoles (multipoles) occurs. Such an orbital rotation of the entire-structure topological magnetic dipoles (multipoles) appears due to geometrical phases on a lateral surface of a ferrite disk at the MDM resonances. It was shown [9, 10] that such topological (geometrical) phases arise from the chiral edge states (chiral Majorana edge states) on a lateral surface of a ferrite disk. The persistent edge currents for magnetization arise because of winding properties essential for the motion of the magnetization in a confined cylindrical geometry. The chiral transport means one-direction propagation of excitations at a given direction of time. In such a structure, a spatial version of the causality principle emerges. This can lead to a situation in which earlier events affect only those future events that occur "downstream". So, in addition to the requirement that future events do not affect the past, one also expects that the downstream events do not affect upstream events even in the future. [14].

MDM oscillations in a quasi-2D ferrite disk can conserve energy and angular momentum [7 – 10]. Because of these properties, MDMs are strongly coupled to microwave fields and enable to confine microwave radiation energy in subwavelength scales. MDM chiral currents strongly modify microwave radiation acting on a ferrite disk. In a vacuum subwavelength region abutting to a MDM ferrite disk, one can observe the quantized-state power-flow vortices [11, 12]. In such a region, a coupling between the time-varying electric and magnetic fields is different from such a coupling in regular electromagnetic (EM) fields. These specific near fields, originated from MDM oscillations, we term magnetoelectric (ME) fields [13]. The ME-field solutions give evidence for spontaneous symmetry breakings at the resonant states of MDM oscillations. In Ref. [15] it was shown that the ME field properties are related to the space-time curvature. Because of rotations of localized field configurations in a fixed observer inertial frame, the linking between the EM and ME fields cause violation of the Lorentz symmetry of spacetime. In such a sense, ME fields can be considered as Lorentz-violating extension of the Maxwell equations [16, 17]. The ME fields have helical (chiral) structure. There are the right-hand (RH) and left-hand (LH) helices [15]. Topological pictures of interaction of ME fields with EM fields well illustrate these helices [15]. If we go through such a chiral structure, we can "worked through" back geometrical phases when changing a direction of a bias magnetic field.

In this paper, we show that due to a topological structure of ME fields one obtains unidirectional multiresonant tunneling for electromagnetic waves propagating in microwave structures with embedded quasi-2D ferrite disks. The resonances manifest themselves as peaks in the scattering-matrix parameters at the stationary states of MDM oscillations in a ferrite disk. The effect of unidirectional tunneling is exhibited as the Fano-resonance interferences, which are different for oppositely directed bias magnetic fields. In 1961, Fano proposed [18] that in a system where a discrete energy level is embedded in a continuum energy state and there is coupling between these two states, a specific resonant state arises around the discrete level. This quantum mechanical interference yields a characteristic asymmetric line shape in the transition probability. Fano-resonance tunneling is a well known effect in semiconductor quantum-well and quantum-dot structures [19, 20]. In Ref. [21] it was shown that interaction of a MDM ferrite particle with its microwave-structure environment has a deep analogy with the Fano-resonance interference in natural and artificial atomic structures. It was stated that MDM particles have the energy-eigenstate spectra and tunneling of microwave radiation into a MDM ferrite disk is due to twisting excitation. All these effects allow us to establish, in the present studies, a close connection between the the observed Fano-resonance tunneling, geometry of a microwave structure, and topology of ME fields. Our observations support the model of a special role of the MDM chiral edge states in the unidirectional ME-field multiresonant tunneling.

The paper is organized as follows. In Section II, we show how such general notions as nonreciprocity and unidectionality are related to gyromagnetic properties of media. We analyze the effects of nonreciprocity and unidirectionality for oscillating modes in resonant structures. In Section III, we analyze chiral edge states of MDMs in a quasi-2D ferrite disk. In Section IV, we represent our numerical and experimental results on the unidirectional ME-field multiresonant tunneling. In Section V we provide some discussions and summarize our studies.

## II. NONRECIPROCITY AND UNIDIRECTIONALITY

The Fano-resonance tunnelling is observed as unidirectional phenomena in microwave structures with MDM ferrite resonators. For deeper understanding these effects, we should dwell initially on such fundamental notions as nonreciprocity and unidectionality in gyrotropic structures. As a general consideration, we analyze the scattering-matrix (*S*-matrix) properties of two-port (input-output) structures with such field behaviors. We also analyze the nonreciprocity and unidirectionality properties for oscillating modes in resonant structures.

## A. Field nonreciprocity and unidirectionality in gyrotropic structures

The effect of nonreciprocity in the electromagnetic wave propagation is observed in gyrotropic media. In classical electromagnetism, Lorentz reciprocity is considered as the most common and general theorem for time-invariant linear media. It involves the interchange of time-harmonic sources and the resulting electromagnetic fields. In isotropic media, reciprocity can be recognized and exploited even in the presence of absorption, whilst time-reversal symmetry precludes absorption. Gyrotropic (gyromagnetic or gyroelectric) media with nonsymmetrical constitutive tensors caused by an applied dc magnetic field have been called *nonreciprocal* media because the usual reciprocity theorem [22] does not apply to them. Rumsey has introduced a quantity called the "reaction" and interpreted it as a "physical observable" [23]. This made it possible to obtain a modified reciprocity theorem based on the property of gyrotropic media that nonsymmetrical constitutive tensors of permittivity or permeability are transposed by reversing the dc magnetic field $\vec{H}_0$ [24]. Based on this aspect, a widely used formulation is the following: devices that are nonreciprocal in their electromagnetic properties are so because of asymmetry of the magnetic or dielectric tensors of the linear media they contain. This concerns both microwave ferrite-based devices and magneto-optic-based devices. For the scattering-matrix parameters in nonreciprocal devices, one has $S_{ij}(\vec{H}_0) = S_{ji}(-\vec{H}_0), \ i \neq j$. For a lossless structure with a gyrotropic medium, the scattering matrix is unitary. On the microscopic level, the applicability of the reciprocity theorem for gyrotropic media is based on the time-reversal invariance, which is described by the Onsager principle [25], [26].

The notion of unidirectionality in gyrotropic structures is different from the notion of nonreciprocity. To suppress backscattering, the structure should be lossy for backscattered propagating waves. So, a scattering matrix should be non-unitary. The *S*-matrix for an ideal two-port isolator (the structure with complete suppression of backscattered energy) has the form $[S] = \begin{bmatrix} 0 & 0 \\ 1 & 0 \end{bmatrix}$, indicating that both ports are matched, but transmission occurs only in direction from post 1 to port 2. While the *S* matrix is not unitary, it is also not symmetric. Working of an ideal isolator with a gyrotropic medium is based on the *nonreciprocal-absorption* properties. In a general case, we have: $\left( \left| S_{ij} \right| (\vec{H}_0) \right) \neq \left( \left| S_{ji} \right| (\vec{H}_0) \right)$, but $\left( \left| S_{ij} \right| (\vec{H}_0) \right) = \left( \left| S_{ij} \right| (-\vec{H}_0) \right)$ [27]. This analysis on unidirectionality makes questionable the statement in [1 − 3] that in *lossless* systems with broken time-reversal symmetry, one-way propagation of electromagnetic-wave energy leads to suppression of backscattering.

## B. Nonreciprocity and unidirectionality for oscillating modes in resonant structures

Some general aspects of nonreciprocity and unidirectionality in gyrotropic structures are well illustrated in consideration of oscillating modes in resonant structures. This consideration will allow better understanding of our analysis of the properties of MDM oscillations in a ferrite-disk resonator.

For the electric and magnetic fields in isotropic media, represented in the wave-number space as $\vec{E} = \int \vec{E}_{\vec{k}} e^{-i\vec{k}\vec{r}} d^3k$ and $\vec{H} = \int \vec{H}_{\vec{k}} e^{-i\vec{k}\vec{r}} d^3k$, the reality of the fields means that

$$\vec{E}_{-\vec{k}} = \vec{E}_{\vec{k}}^* \qquad \text{and} \qquad \vec{H}_{-\vec{k}} = \vec{H}_{\vec{k}}^* \ . \qquad (1)$$

This makes it possible to obtain the orthonormality conditions for electromagnetic waves of oscillating modes in a lossless distributed-parameter resonator with an isotropic medium:

$$\int_V \left( \vec{E}_{-\vec{k}} \right)_m \left( \vec{E}_{\vec{k}} \right)_n dV = \int_V \left( \vec{E}_{\vec{k}} \right)_m^* \left( \vec{E}_{\vec{k}} \right)_n dV = \delta_{mn},$$

$$\int_V \left( \vec{H}_{-\vec{k}} \right)_m \left( \vec{H}_{\vec{k}} \right)_n dV = \int_V \left( \vec{H}_{\vec{k}} \right)_m^* \left( \vec{H}_{\vec{k}} \right)_n dV = \delta_{mn}, \qquad (2)$$

where $\delta_{mn}$ is the Kronecker delta. Eqs. (2) show that in such a resonator one can normalize the field of a given mode or to the field of a corresponding counter propagating mode, or to the complex conjugated field of the same mode. Let us dwell now on a lossless running-wave ring resonator containing an isotropic medium. In this case, the resonance occurs due to a wave running only in one direction along a circle. Nevertheless, also in such a resonator the orthonormality conditions, expressed by Eq. (2), take place since for every clockwise running mode one may have the same-type counterclockwise running mode, and vice versa. It means that for rotary motion of energy in one direction there exists the counterpart – the rotary motion of energy in the opposite direction. When an input monochromatic signal is not exactly at the resonance frequency, one has a reflected wave. Now the following question arises: Can we realize a lossless running-wave ring resonator based on a one-way waveguide with unidirectional propagation of electromagnetic energy? More generally speaking, can one create an electromagnetic-wave resonator with the fields nonsymmetrical in the $\vec{k}$ space? Suppose that based on any of the structures described in Refs. [1 – 3] we realized a large in-plane loop. A radius of this loop is much bigger than the wavelength of electromagnetic waves. So, locally, one has a one-way waveguide with the properties shown in Refs. [1 – 3]. It is evident that in a case of one-wave propagation of energy, a small fluctuation of the field energy in a certain point of the loop will lead to infinite accumulation of energy in an entire lossless structure during multi-cyclic processes of the wave propagation. Such a rotary motion of energy in only one direction, while preventing motion in the opposite direction (a ratchet device), is beyond the laws of thermodynamics. Energy of a signal cannot propagate only in one direction in such a travelling-wave loop. The proposed structure must be lossy. In other words, to realize a waveguide with one-wave propagation of energy, one has to create certain channels for energy losses. In such a case, the fields are not real and so the orthonormality conditions, expressed by Eq. (2), cannot be fulfilled.

While realization of a lossless resonator with unidirectional propagation of electromagnetic energy is, physically, a meaningless problem, the question is about possibility to create a lossless gyromagnetic electromagnetic-wave resonator with nonreciprocal wave propagation and real fields. When a lossless resonator contains a gyromagnetic medium characterized by a permeability tensor $\bar{\bar{\mu}}$, the orthonormality conditions for electromagnetic waves are expressed as [28]

$$\int_V \left(\vec{E}_{\vec{k}}\right)_m^* \left(\vec{E}_{\vec{k}}\right)_n dV = \delta_{mn}, \qquad \int_V \left(\vec{H}_{\vec{k}}\right)_m^* \left[\vec{\mu}\left(\vec{H}_{\vec{k}}\right)_n\right] dV = \delta_{mn}. \tag{3}$$

The tensor $\vec{\mu}$ is a Hermitian tensor:

$$\left(\vec{\mu}\right)^T = \left(\vec{\mu}\right)^*. \tag{4}$$

Moreover, from the Onsager relations for kinetic coefficients, one has for the components of the tensor $\vec{\mu}$ [25, 26, 28]:

$$\mu_{ij}\left(\omega, \vec{H}_0\right) = \mu_{ji}\left(\omega, -\vec{H}_0\right), \tag{5}$$

where $\vec{H}_0$ is an external bias magnetic field.

The orthonormality conditions (3) are different from the orthonormality conditions (2). A general analysis shows that an electromagnetic-wave resonator containing a gyromagnetic-medium sample has real eigenfrequencies, but complex eigenfunctions [28]. It means that there is no a standing-wave behavior of the resonator eigenfunctions. Let us consider, for example, a ferrite cylinder of the length $l$ longitudinally magnetized in direction of axis $z$. If one uses separation of variables, at the resonance frequency there will be a standing wave along $z$-axis and, because of the Faraday rotation, an azimuthally running wave in the plane perpendicular to $z$-axis. At resonance frequency, the phases of the waves along $z$-axis and in the plane perpendicular to $z$-axis should be correlated. But the phase of the azimuthally running wave is not identified in the cross-sectional plane of a ferrite cylinder. So, one should conclude that such a representation of the eidenfunction as a standing-wave plus running-wave behavior is incorrect. In a lossless gyromagnetic electromagnetic-wave resonator with nonreciprocal wave propagation one cannot identify a certain phase difference between two given points at a resonance frequency.

The main reason for this is that microwave resonators with ferrite inclusions are nonintegrable systems because of the time-reversal symmetry (TRS) breaking effects. The concept of nonintegrable, i.e. path-dependent, phase factors is one of the fundamental concepts of electromagnetism. A key aspect of the behavior of the ferrite-resonator configuration concerns reflection and refraction of electromagnetic waves at ferrite-vacuum interfaces. In a general case of oblique incidence of a wave on a single ferrite-vacuum interface, apparently different situations arise by changing the directions of incident waves and bias, and incident side of the interface [29 – 31]. In a system of a cavity and a ferrite sample one obtains a TRS-breaking microwave billiard. Due to the TRS breaking, the ferrite-vacuum boundary conditions entail the existence of the solutions to the differential equations which are topologically distinct. This leads to creation of topological defects – the Poynting-vector vortices [30 – 33].

Can one create a microwave ferrite resonator with real eigenstates and real eigenfunctions? Yes. But there will be the resonator with magnetostatic-wave (MS-wave) [or the magnetic-dipolar-mode (MDM)] oscillations. It is well known that in a case of small (compared to the free-space electromagnetic-wave wavelength) samples made of magnetic media with strong temporal dispersion, the role of an electric displacement current in Maxwell equations can be negligibly small, so oscillating fields are the quasistationary fields [26]. A magnetic field $\vec{H}$ is a quasimagnetostatic field ($\vec{\nabla} \times \vec{H} = 0$), which is expressed by a magnetostatic potential: $\vec{H} = -\nabla \psi$. The spectral properties of oscillations in such a small ferrite sample are analyzed based on the Walker equation for MS-potential wave function $\psi(\vec{r}, t)$ [34]:

$$\vec{\nabla} \cdot \left( \vec{\mu} \cdot \vec{\nabla} \psi \right) = 0 \,. \tag{6}$$

Outside a ferrite this equation becomes the Laplace equation. The MDM oscillations appear because of a prevailing role of long-range dipole-dipole interactions in a small ferrite sample. Importantly, excitation of the real-eigenstate multiresonance MDM oscillations in a quasi-2D ferrite disk by microwave radiation, observed, for the first time, in Ref. [35], is possible due to presence of surface chiral currents. In the near-field region, these chiral currents create specific topologically distinctive structures – the ME-fields – resulting in observation of the unidirectional multiresonant tunneling, discussed in this paper.

## III. MDM CHIRAL CURRENTS AND THEIR INTERACTION WITH MICROWAVE RADIATION

### A. MDM eigenvalue problems, chiral states, and quantized electric fluxes

MDM oscillations in a quasi-2D ferrite disk are mesoscopically quantized states. Long range dipole-dipole correlation in position of electron spins in a ferrimagnetic sample with saturation magnetization can be treated in terms of collective excitations of the system as a whole. If the sample is sufficiently small so that the dephasing length $L_{ph}$ of the magnetic dipole-dipole interaction exceeds the sample size, this interaction is non-local on the scale of $L_{ph}$. This is a feature of a mesoscopic ferrite sample, i.e., a sample with linear dimensions smaller than $L_{ph}$ but still much larger than the exchange-interaction scales. In a case of a quasi-2D ferrite disk, the quantized forms of these collective matter oscillations – magnetostatic magnons – were found to be quasiparticles with both wave-like and particle-like behavior, as expected for quantum excitations. The magnon motion in this system is quantized in the direction perpendicular to the plane of a ferrite disk. The MDM oscillations in a quasi-2D ferrite disk, analyzed as spectral solutions for the MS-potential wave function $\psi(\vec{r},t)$, has evident quantum-like attributes [7 – 10].

In microwave experiments with a normally magnetized quasi-2D ferrite disk, regular multiresonance MDM spectra have been observed [21, 35 – 39]. It was shown that in such a ferrite-disk resonator, MDM oscillations can be characterized by real eigenstates and real eigenfunctions. Formulation of quasi-Hermitian eigenvalue problem and analytical spectral solutions for MDMs in a normally magnetized ferrite disk were obtained in Ref. [7, 8]. For the disk geometry, the energy-eigenstate oscillations are described by a two-dimensional (with respect to in-plane coordinates of a disk) differential operator $\hat{G}$:

$$\hat{G}_{\perp} = \frac{g_q}{16\pi} \mu \, \nabla_{\perp}^2 \,, \tag{7}$$

where $\nabla_{\perp}^2$ is the two-dimensional Laplace operator, $\mu$ is a diagonal component of the permeability tensor, and $g_q$ is a dimensional normalization coefficient for mode $q$. Operator $\hat{G}_{\perp}$ is positive definite for negative quantities $\mu$. The normalized average (on the RF period) density of accumulated magnetic energy of mode $q$ is determined as

$$E_q = \frac{g_q}{16\pi}\left(\beta_{z_q}\right)^2, \tag{8}$$

where $\beta_{z_q}$ is the propagation constant of mode $q$ along the disk axis $z$. The energy eigenvalue problem is defined by the differential equation:

$$\hat{G}_\perp \tilde{\eta}_q = E_q \tilde{\eta}_q, \tag{9}$$

where $\tilde{\eta}_q$ is a dimensionless membrane ("in-plane") MS-potential wave function. In the energetic representation, a square of a modulus of the wave function defines probability to find a system with a certain energy value. The scalar-wave membrane function $\tilde{\eta}$ can be represented as

$$\tilde{\eta} = \sum_q a_q \tilde{\eta}_q \tag{11}$$

and the probability to find a system in a certain state $q$ is defined as

$$|a_q|^2 = \left|\int_S \tilde{\eta}\ \tilde{\eta}_q^* dS\right|^2. \tag{12}$$

MDM oscillations in a ferrite disk are described by real eigenfunctions: $\tilde{\eta}_{-\tilde{\beta}} = \tilde{\eta}_{\tilde{\beta}}^*$. The orthonormality conditions are expressed as

$$\int_S \left(\tilde{\eta}_{\tilde{\beta}}\right)_q \left(\tilde{\eta}_{-\tilde{\beta}}\right)_{q'} dS = \int_S \tilde{\eta}_q \tilde{\eta}_{q'}^* dS = \delta_{qq'}, \tag{13}$$

where $S$ is a cylindrical cross section of a ferrite disk.

The above spectral solutions, based on differential operator $G$, we conventionally call $G$-mode solutions. In solving the energy-eigenstate spectral problem for the $G$-mode states, the boundary condition on a lateral surface of a ferrite disk, is expressed as [6 − 8]

$$\mu\left(\frac{\partial \tilde{\eta}}{\partial r}\right)_{r=\Re^-} - \left(\frac{\partial \tilde{\eta}}{\partial r}\right)_{r=\Re^+} = 0, \tag{14}$$

where $\Re$ is a radius of a ferrite disk. There is a homogeneous boundary condition for a differential operator $\hat{G}_\perp$ [see Eq. (7)]. For the magnetic field components, Eq. (14) is written as

$$\mu(H_r)_{r=\Re^-} - (H_r)_{r=\Re^+} = 0, \tag{15}$$

where $(H_r)_{r=\Re^-}$ and $(H_r)_{r=\Re^+}$ are the radial components of a magnetic field on a border circle of a ferrite disk.

The $G$-mode ferrite disk, which is not connected to the surrounding, returns to the original situation after $2\pi$ rotation. The $G$-mode object connected to the microwave surrounding behaves differently. It appears that the $G$-mode boundary conditions are different from standard electromagnetic boundary conditions. The $G$-mode spectrum is obtained based on solution of the Walker equation (6). This equation is, in fact, the magnetostatic-description representation of a differential equation $\nabla \cdot \vec{B} = 0$. It is evident that the boundary condition (15) manifests itself in contradictions with the boundary condition for continuity of a radial component of magnetic flux density $\vec{B}$ on a lateral surface of a ferrite-disk resonator. Such a boundary condition should be written as

$$\mu (H_r)_{r=\Re^-} - (H_r)_{r=\Re^+} = -i\mu_a (H_\theta)_{r=\Re} , \qquad (16)$$

where $(H_\theta)_{r=\Re}$ is an annular magnetic field on a border circle. In the MS description, this equation appears as

$$\mu \left( \frac{\partial \tilde{\varphi}}{\partial r} \right)_{r=\Re^-} - \left( \frac{\partial \tilde{\varphi}}{\partial r} \right)_{r=\Re^+} = -\mu_a \nu \left( \tilde{\varphi} \right)_{r=\Re^-} , \qquad (17)$$

where $\tilde{\varphi}$ is the MS-potential membrane wave function, $\nu$ is an azimuth wave number, and $\mu_a$ is a off-diagonal component of the permeability tensor. Contrary to real wave functions $\tilde{\eta}$, functions $\tilde{\varphi}$ are complex wave functions. The term in the right-hand side of Eq. (17) has the off-diagonal component of the permeability tensor, $\mu_a$, in the first degree. There is also the first-order derivative of function $\tilde{\varphi}$ with respect to the azimuth coordinate. It means that for the MS-wave solutions one can distinguish the time direction (given by the direction of the magnetization precession and correlated with a sign of $\mu_a$) and the azimuth rotation direction (given by a sign of $\partial \tilde{\varphi} / \partial \theta$). For a given sign of a parameter $\mu_a$, there are different MS-potential wave functions, $\tilde{\varphi}^{(+)}$ and $\tilde{\varphi}^{(-)}$, corresponding to the positive and negative directions of the phase variations with respect to a given direction of azimuth coordinates, when $0 \leq \theta \leq 2\pi$. There is an evidence for the path dependence in the problem solutions.

The $G$-mode solutions obtained based on the boundary condition (15) are the stationary-state solutions with singlevalued MS-potential wave functions $\tilde{\eta}$. Contrary, the spectral solutions obtained based on the boundary condition (16) cannot be considered as stationary-state solutions with singlevalued MS-potential wave functions. A singular border term in the right-hand-side of Eq. (16), which expresses the discontinuity of a radial component of magnetic flux density for the $G$-mode solutions, we represent as effective surface magnetic charge density:

$$-i\mu_a \left( H_\theta \right)_{r=\Re} \equiv 4\pi \rho_s^{(m)} . \qquad (18)$$

This equation relates an azimuthal component of the magnetic field with surface magnetic charge density. In fact, the charges $\rho_s^{(m)}$ are topological magnetic charges.

We can consider surface magnetic charge density $\rho_s^{(m)}$ as a certain fluctuation. Such magnetic-charge edge states of MDMs contribute to appearance of a surface magnetic current around the border ring. For the time varying $G$-mode fields, the surface magnetic charge density $\rho_s^{(m)}$ should be related to the surface magnetic current density $\vec{j}_s^{(m)}$ by a continuity equation. For monochromatic wave process ($\sim e^{i\omega t}$), we have:

$$\vec{\nabla} \cdot \vec{j}_s^{(m)} = -i\omega \rho_s^{(m)} . \qquad (19)$$

Both quantities, $\rho_s^{(m)}$ and $\vec{j}_s^{(m)}$, have time- and space-dependent phases. Evidently, magnetic charges $\rho_s^{(m)}$ appear in a form of the magnetic-dipole (magnetic-multipole, in general) structure on a lateral surface of a ferrite disk. Magnetic currents $\vec{j}_s^{(m)}$ are not linear, but circulating currents. So, Eq. (19) can take place only when magnetic charges $\rho_s^{(m)}$ are the charges moving (clockwise or counterclockwise) on a lateral surface of a ferrite disk. This gives evidence for the fact that the $G$-mode picture should rotate in the laboratory frame when the magnetic charges $\rho_s^{(m)}$ exist. Definitely, surface magnetic charges $\rho_s^{(m)}$ are not "free magnetic charges". There are topological charges determined by orientation of the magnetization vectors on a lateral surface of a ferrite disk. Also, surface magnetic currents $\vec{j}_s^{(m)}$ are not caused by motion of "free magnetic charges". In fact, there are chiral-rotation surface magnetostatic spin waves.

In Fig. 1, we show surface magnetic charges and edge chiral magnetic currents. This is a view on the upper plane of a ferrite disk. Suppose that there exists a magnetic-dipole fluctuation on a lateral surface of a ferrite disk and a surface magnetic current is a clockwise rotating surface wave. Let at a given time moment, there be a positive magnetic charge $(+)^{(m)}$ at point $P$ on a disk lateral surface and a negative magnetic charge $(-)^{(m)}$ at a diametrically opposite point $Q$. Let there be a magnetic current $\left[ \vec{j}_s^{(m)} \right]_A$ "departing", with a certain phase $\tau_A$, from a point $P$. Because of conservation of "magnetic neutrality", another current wave $\left[ \vec{j}_s^{(m)} \right]_B$ with the same phase $\tau_B = \tau_A$ should "depart" to a point $Q$. Since, however, no real magnetic charges physically exist, these magnetic currents on a lateral surface of a ferrite disk can be only the topological currents. Topologically, a circulating current $\left[ \vec{j}_s^{(m)} \right]_A$ is not the same as a circulating current $\left[ \vec{j}_s^{(m)} \right]_B$. Every of these separate currents gets around the orbital trajectory $\mathcal{L} = 2\pi\Re$ during a half of period of microwave radiation. The current $\left[ \vec{j}_s^{(m)} \right]_A$ "arrives" to a point $P$ with the phase $-\tau_A$ when the $G$-mode rotates at the angle of $2\pi$. The similar situation is for the current $\left[ \vec{j}_s^{(m)} \right]_B$. So, we can state that for the $G$-mode regular-coordinate angle $\theta = 2\pi$, a topological surface magnetic current acquires the phase of $\theta' = \pi$. Because of a magnetic-dipole fluctuation on a lateral surface of a ferrite disk the domain of the $G$-mode azimuthal angle $\theta$, in a laboratory frame, is no more $[0, 2\pi]$ but $[0, 4\pi]$.

Circulating currents $\left[ \vec{j}_s^{(m)} \right]_A$ and $\left[ \vec{j}_s^{(m)} \right]_B$ appear as topologically distinctive currents due to finite thickness of the disk. Fig. 2 illustrates a two-layer-ring model for surface magnetic currents. When a magnetic current of the upper (lower) layer is arriving to terminal $P$ (where a topological magnetic charge is supposed to be localized), it must continue its propagation at the lower (upper) layer and this is only one choice. The similar situation takes place at terminal $Q$. Regions of terminals $P$ and $Q$ are the regions of singularity. At the same time, we have to note that topological magnetic charges are not point charges. They are distributed on a lateral surface of the disk. So, sharp "kinks" of the current lines are impossible.

Due to the special topology of the two-layer ring, orbital angular momenta are allowed to be a half-integer. In a quasi-2D ferrite disk, the two layers are very close to each other and the above two currents look like a ring magnetic current on a lateral surface of a ferrite disk. So, the continuity equation (19) has a form:

$$\vec{\nabla}_\theta \cdot \left( \vec{j}_s^{(m)} \right)_\theta = -i\omega \rho_s^{(m)}, \tag{20}$$

where $\vec{\nabla}_\theta \cdot \left( \vec{j}_s^{(m)} \right)_\theta = \dfrac{1}{\Re} \dfrac{\partial \left( \vec{j}_s^{(m)} \right)_\theta}{\partial \theta}$. The closed-loop surface magnetic current $\left( \vec{j}_s^{(m)} \right)_\theta$ are clockwise and counterclockwise edge chiral currents depending on a direction of a bias magnetic field (in other words, depending on a direction of time).

In a quasi-2D ferrite disk, the edge chiral currents $\left( \vec{j}_s^{(m)} \right)_\theta$ can be described by 1D scalar wave functions, which are double-valued functions. Such ring magnetic currents can create electric-field fluxes. This effect was studied in details in Ref. [9, 10, 13]. On a lateral border of a ferrite disk, the correspondence between a double-valued membrane wave function $\tilde{\varphi}$ and a singlevalued membrane function $\tilde{\eta}$ is expressed as: $\left( \tilde{\varphi}_\pm \right)_{r=\Re} = \delta_\pm \left( \tilde{\eta} \right)_{r=\Re^-}$, where $\delta_\pm \equiv f_\pm e^{-iq_\pm \theta}$ is a double-valued edge wave function on contour $\mathcal{L} = 2\pi\Re$. The azimuth number $q_\pm$ is equal to $\pm \dfrac{1}{2} l$, where $l$ is an odd quantity ($l = 1, 3, 5, \dots$). For amplitudes we have $f_+ = -f_-$ and $|f_\pm| = 1$. Function $\delta_\pm$ changes its sign when $\theta$ is rotated by $2\pi$ so that $e^{-iq_\pm 2\pi} = -1$. As a result, one has the eigenstate spectrum of MDM oscillations with topological phases accumulated by the edge wave function $\delta$. On a lateral surface of a quasi-2D ferrite disk, one can distinguish two different functions $\delta_\pm$, which are the counterclockwise and clockwise rotating-wave edge functions with respect to a membrane function $\tilde{\eta}$. A line integral around a singular contour $\mathcal{L}$:

$\dfrac{1}{\Re} \oint_\mathcal{L} \left( i \dfrac{\partial \delta_\pm}{\partial \theta} \right) (\delta_\pm)^* \, d\mathcal{L} = \displaystyle\int_0^{2\pi} \left[ \left( i \dfrac{\partial \delta_\pm}{\partial \theta} \right) (\delta_\pm)^* \right]_{r=\Re} d\theta$ is an observable quantity. It follows from the fact that because of such a quantity one can restore singlevaluedness of the spectral problem. Because of the existing the geometrical phase factor on a lateral boundary of a ferrite disk, MDMs are characterized by a pseudo-electric field (the gauge field) [9, 13]. We will denote here this pseudo-electric field by the letter $\vec{\mathcal{E}}$. The geometrical phase factor in the $G$-mode solution is not single-valued under continuation around a contour $\mathcal{L}$ and can be correlated with a certain vector potential $\vec{\Lambda}_\mathcal{E}^{(m)}$. We define a geometrical phase for a MDM as [9, 13]

$$i\Re \int_0^{2\pi} [(\vec{\nabla}_\theta \delta_\pm)(\delta_\pm)^*]_{r=\Re} d\theta \equiv K \oint_\mathcal{L} \left( \vec{\Lambda}_\mathcal{E}^{(m)} \right)_\pm \cdot d\vec{\mathcal{L}} = 2\pi q_\pm. \tag{21}$$

where $\vec{\nabla}_\theta \delta_\pm = \dfrac{1}{\Re} \dfrac{\partial \delta_\pm}{\partial \theta}\bigg|_{r=\Re} \vec{e}_\theta$ and $\vec{e}_\theta$ is a unit vector along an azimuth coordinate. In Eq. (21), $K$ is a normalization coefficient. The physical meaning of coefficient $K$ was discussed in Refs [9, 13, 40]. In Eq. (21) we inserted a connection which is an analogue of the Berry phase. In our case, the Berry's phase is generated from the broken dynamical symmetry. The confinement effect for magnetic-dipolar oscillations requires proper phase relationships to guarantee single-valuedness of the wave functions. To compensate for sign ambiguities and thus to make wave functions single-valued we added a vector-potential-type term $\vec{\Lambda}_\mathcal{E}^{(m)}$ (the Berry connection) to the MS-potential Hamiltonian. On a singular contour $\mathcal{L} = 2\pi\Re$, the vector potential $\vec{\Lambda}_\mathcal{E}^{(m)}$ is related to

double-valued functions. It can be observable only via the circulation integral over contour $\mathcal{L}$, not pointwise. The pseudo-electric field $\vec{\epsilon}$ can be found as

$$\vec{\epsilon}_\pm = -\vec{\nabla} \times \left( \vec{\Lambda}_\epsilon^{(m)} \right)_\pm. \tag{22}$$

The field $\vec{\epsilon}$ is the Berry curvature. In contrast to the Berry connection $\vec{\Lambda}_\epsilon^{(m)}$, which is physical only after integrating around a closed path, the Berry curvature $\vec{\epsilon}$ is a gauge-invariant local manifestation of the geometric properties of the MS-potential wavefunctions. The corresponding flux of the gauge field $\vec{\epsilon}$ through a circle of radius $\mathfrak{R}$ is obtained as:

$$K \int_S \left( \vec{\epsilon} \right)_\pm \cdot d\vec{S} = K \oint_\mathcal{L} \left( \vec{\Lambda}_\epsilon^{(m)} \right)_\pm \cdot d\vec{\mathcal{L}} = K \left( \Xi^{(e)} \right)_\pm = 2\pi q_\pm, \tag{23}$$

where $\left( \Xi^{(e)} \right)_\pm$ are quantized fluxes of pseudo-electric fields. Each MDM is quantized to a quantum of an emergent electric flux. There are the positive and negative eigenfluxes. These different-sign fluxes should be nonequivalent to avoid the cancellation. It is evident that while integration of the Berry curvature over the regular-coordinate angle $\theta$ is quantized in units of $2\pi$, integration over the spin-coordinate angle $\theta'$ $\left( \theta' = \frac{1}{2}\theta \right)$ is quantized in units of $\pi$. The physical meaning of coefficient $K$ in Eqs. (21), (23) concerns the property of a flux of a pseudo-electric field.

The edge chiral current $\vec{j}_s^{(m)}$ is a persistent magnetic current in an Aharonov-Bohm-like geometry. On an edge ring region, we have the magnetization motion pierced by an electric flux $\left( \Xi^e \right)_\pm$. The edge magnetic current can be observable only via its circulation integrals, not pointwise. This results in the moment oriented along a disk normal. It was shown experimentally [37] that such a moment has a response in an external RF electric field and so can be classified as an electric moment. There is a so called anapole moment $a_\pm^e$ [9].

**B. Interaction of MDMs with external microwave radiation**

Because of the edge chiral currents, interaction of MDMs with external microwave radiation of MDM oscillations is manifested by unique topological properties.

In numerous microwave experiments with a quasi-2D ferrite disk, multiresonant MDM oscillations were observed both at a constant signal frequency with scanning of a bias magnetic field and at a constant bias magnetic field with scanning of a signal frequency [21, 35 – 39]. In initial experimental studies [35, 36], it was shown that in a microwave structure with an embedded quasi-2D ferrite disk, multiresonance MDM oscillations are excited by RF magnetic fields lying in the disk plane. Later, it was shown experimentally that MDMs can also be effectively excited by RF electric field of a microwave structure, which is oriented along a disk axis [37]. One of the main conclusions, we can made from all these experiments, is the fact that positions of the resonance peaks are not dependent on the microwave-structure environment and are exceptionally determined by the disk parameters. It means that in all the experiments we observe the energy-eigenstate spectra. As we discussed above, the domain of the G-mode azimuthal angle $\theta$ is not $[0, 2\pi]$ but $[0, 4\pi]$. The G-mode object, being connected to the microwave surrounding, returns to the original situation after $4\pi$ rotation. It means that for a given frequency $\omega$ of microwave radiation, the $G$-mode rotation frequency is $2\omega$.

When (due to interaction with external microwave radiation) a macroscopic MS-potential wave function $\tilde{\eta}$ is set into rotation, quantized vortex lines appears. At the vortex center (the center of a ferrite disk) the function $\tilde{\eta}$ is zero. In different physical systems, there are many other examples of such vortices with rotating macroscopic wave functions. In particular, there are vortices in Bose-Einstein-condensate systems [41 – 44]. The operator $\hat{G}_\perp$ in Eq. (7) is a two-dimensional differential operator in the absence of rotation. The eigenstate of this operator is the energy of the ground state with no vortices. In a rotating frame, this differential operator has a form:

$$\hat{G}'_\perp = \hat{G}_\perp - \Omega \hat{L}_z . \qquad (24)$$

The last term in the right-hand side of this equation favors of states with non-zero angular momentum. $\hat{L}_z = i\left( y \dfrac{\partial}{\partial x} - x \dfrac{\partial}{\partial y} \right)$ is the orbital angular momentum along the rotation axis $z$ – the disk axis. A rotation frequency is $\Omega = 2\omega$. For a certain MDM $q$, the energy of a vortex state in a frame rotating with angular frequency $\Omega_q$ is $E'_q = E_q - \Omega_q \hat{L}_{z_q}$.

Microwave radiation can potentially couple to MDM oscillations if the ferrite sample shows a confined structure to satisfy conservation of energy and angular momentum. Tunneling of microwave radiation into a MDM ferrite disk is due to twisting excitations. Our studies give evidence for such near-field twisting excitations. There are subwavelength field structures with quantized energy and angular momentums. It is known that photons, like other particles, carry energy and angular momentum. A circularly polarized photon carries a spin angular momentum [45]. Also, photons can carry additional angular momentum, called orbital angular momentum. Such photons, carrying both spin and orbital angular momentums are called twisting photons [46]. Twisting photons are propagating-wave behaviors. These are "real photons". In the near-field phenomena, which have subwavelength-range effects and do not radiate through space with the same range-properties as do electromagnetic wave photons, the energy is carried by virtual photons, not actual photons. Virtual particles should also conserve energy and momentum. The question whether virtual photons can behave as twisting excitations, is a subject of numerous discussions [47 – 50]. In particular, in Refs. [47, 49] it was discussed that vacuum can induce a torque between two uniaxial birefringent dielectric plates. In this case, the fluctuating electromagnetic fields have boundary conditions that depend on the relative orientation of the optical axes of the materials. Hence, the zero-point energy arising from these fields also has an angular dependence. This leads to a Casimir torque that tends to align two of the principal axes of the material in order to minimize the system's energy. A torque occurs only if symmetry between the right-hand and left-hand circularly polarized light is broken (when the media are birefringent). In our case, a quantum vacuum field takes energy from the MDM ferrite disk. The electromagnetic mode with frequency $\omega$ interacts with the MDM of frequency $2\omega$ (because the ac part of the ponderomotive force has frequency $2\omega$). This is the parametric pumping of the energy from magnetomechanical (magnetization-precession) oscillations into electromagnetic oscillations. Energy taken is converted into real photons. This is the dynamical Casimir effect. The dynamical Casimir effect is the generation of photons out of the quantum vacuum induced by an accelerated body. Rotating MDMs in a ferrite disk cannot rule out a dynamical Casimir torque even in the case of uniform angular velocity. This raises the question of the angular-momentum coupling with the quantum vacuum field. Quantized vortices are sensitive probes of the angular-momentum coupling of MDMs with the vacuum field.

The properties of rotating *G* modes and edge chiral states in a ferrite disk essentially determine a character of interaction of MDMs with external microwave radiation. Suppose that a small ferrite disk is placed in a region of a linearly polarized (in a plane of a ferrite disk) RF magnetic field. When no MDM oscillations are assumed, we can state that such a RF magnetic field excites *homogeneous-ferromagnetic-resonance* oscillations of magnetization in a small ferrite sample [28]. One observes an induced magnetic dipole lying in the disk plane. The quantity of this dipole $\vec{P}^{(m)}$ is determined both by the incident-wave magnetic field $\vec{H}^{(i)}$ and by the scattered-wave magnetic field $\vec{H}^{(s)}$:

$$P^{(m)} = \chi^{(e)} \left( \vec{H}^{(i)} + \vec{H}^{(s)} \right), \tag{25}$$

where $\chi^{(e)}$ is the ellipsoid external susceptibility [28]. Formally, one can describe the magnetic-charge distribution on a lateral surface of a ferrite disk as an azimuth magnetic-charge standing wave. There are two equal-amplitude azimuth waves of surface magnetic charges. Also, on a lateral surface of a ferrite disk, the azimuth component of a magnetic field $\vec{H}_{\theta}^{(ext)} = \vec{H}_{\theta}^{(i)} + \vec{H}_{\theta}^{(s)}$ can be represented as an azimuth magnetic-field standing wave. It means that there are two equal-amplitude azimuth waves of a magnetic field: $\vec{H}_{\theta}^{(ext)} = \frac{1}{2} \left( \vec{H}_{+\theta}^{(ext)} + \vec{H}_{-\theta}^{(ext)} \right)$.

At certain quantization conditions, long range dipole-dipole correlation in position of electron spins in a ferrite disk results in appearance of collective excitations of the system. These quantized excitations are MDMs. Surface magnetic charges induced by a linearly polarized (in a plane of a ferrite disk) RF magnetic field can be considered as the sources for MDM oscillations. In this case, however, the magnetic-charge distribution on a lateral surface of a ferrite disk is not an azimuth magnetic-charge standing wave. There exists only a clockwise or only counterclockwise (depending on a direction of a bias magnetic field) azimuth wave of surface magnetic charges. For excitation of *G*-modes by external RF magnetic fields, the right-hand-side term in Eq. (16) appears as an external source. Based on a perturbation-theory analysis, we can assume that the azimuth component of a time-varying incident magnetic field $\left( H_{+\theta} \right)_{r=\Re}^{(i)}$ or $\left( H_{-\theta} \right)_{r=\Re}^{(i)}$ induces surface magnetic charges:

$$\mp i \mu_a \left( H_{\pm\theta} \right)_{r=\Re}^{(i)} \equiv 4\pi \left[ \rho_s^{(m)} \right]_{\pm}^{(i)} \tag{26}$$

On the other hand, with the energy eigenstate description based on the singlevalued *G*-mode MS-potential wave functions $\tilde{\eta}$, one has to show that there exists a certain internal mechanism which removes the perturbation term in a right-hand side of Eq. (16). From the above analysis, we can state that the rotating surface magnetic charge $\left[ \rho_s^{(m)} \right]_{\pm}^{(i)}$ creates, on a lateral surface of a ferrite disk, the surface magnetic current $\left[ \left( \vec{j}_s^{(m)} \right)_{\theta'} \right]^{(i)}$ and this edge chiral current, in its turn, originates an incident-wave electric flux $\left( \Xi^{(e)} \right)^{(i)}$. The *G*-mode stationary-state solution will take place when a certain electric flux appears to *compensate* the incident-wave electric flux $\left( \Xi^{(e)} \right)^{(i)}$. We call this flux as the scattered-wave electric flux $\left( \Xi^{(e)} \right)^{(s)}$. We can characterize this effect as the electric self-inductance effect. The flux $\left(\Xi^{(e)}\right)^{(s)}$ will induce, on a lateral surface of a ferrite disk, the edge magnetic current $\left[\left(\vec{j}_s^{(m)}\right)_{\theta'}\right]^{(s)}$ and this chiral current, in its turn, will create surface magnetic charge $\left[\rho_s^{(m)}\right]^{(s)}$ compensating the incident-wave magnetic charge $\left[\rho_s^{(m)}\right]^{(i)}$. Every of these magnetic currents, $\left[\left(\vec{j}_s^{(m)}\right)_{\theta'}\right]^{(i)}$ and $\left[\left(\vec{j}_s^{(m)}\right)_{\theta'}\right]^{(s)}$, is composed by two topologically distinctive current components discussed above. Both currents, $\left[\left(\vec{j}_s^{(m)}\right)_{\theta'}\right]^{(i)}$ and $\left[\left(\vec{j}_s^{(m)}\right)_{\theta'}\right]^{(s)}$, have the same direction along the azimuth coordinate and are mutually time-phase shifted at $180^{\circ}$. Fig. 3 shows the edge magnetic currents $\left[\left(\vec{j}_s^{(m)}\right)_{\theta'}\right]^{(i)}$ and $\left[\left(\vec{j}_s^{(m)}\right)_{\theta'}\right]^{(s)}$ on a lateral surface of a ferrite disk and their possible correlation with the $G$-mode MS-potential wave functions $\tilde{\eta}$ of a certain MDM.

What will happen when a ferrite disk is placed in the RF electric field oriented normally to a ferrite disk? Following the above model of the electric self-inductance effect, one can state that in this case the incident-wave electric flux $\left(\Xi^{(e)}\right)^{(i)}$ should be compensated by the scattered-wave electric flux $\left(\Xi^{(e)}\right)^{(s)}$. A numerical analysis of the electric field structure clearly proves this statement. In Refs. [11, 50] it was shown that near a MDM-resonant ferrite disk a normal component of the RF electric field is zero during all the time period of microwave radiation.

Physically, the above formal introduction of the quantized fluxes of pseudo-electric fields $\left(\Xi^{(e)}\right)_{\pm}$ can be well justifies based on the Thomas precession effect. Rotating surface-charge magnetic dipoles and edge chiral magnetic currents on a lateral surface of a ferrite disk interact with the $G$-mode oscillations. The underlying physical mechanism is the spin-orbit interaction, which couples the magnetization spin degree of freedom to their orbital dynamics. In Fig. 3, such spin-orbit interaction is illustrated by highlighted parts of in the graphs of the edge magnetic currents $\left[\left(\vec{j}_s^{(m)}\right)_{\theta'}\right]^{(i)}$ and $\left[\left(\vec{j}_s^{(m)}\right)_{\theta'}\right]^{(s)}$ and shaded areas on the graph of the $G$-mode MS-potential wave functions $\tilde{\eta}$. Because of the spin-orbit interaction, the spin is directly related to the winding of trajectory. Such degrees of freedom as geometric (Berry) phases are winding dependent. The Thomas precession talks us about rotation of a spinning particle under the coordinate transformations. The circular spin current along a ring will inevitably produce the Thompson precession. Due to this precession one has an effective electric field. There is the gauge field, no gradient of an electrical potential takes place in this case [51]. Importantly, the electric field due to Thomas precession may exist in any magnetic particle which confines magnetic moments in motion [51].

Rotating surface-charge magnetic dipoles and edge chiral magnetic currents strongly transform both magnetization dynamics inside a ferrite-disk particle and topology of the near-field microwave radiation. As a result of the spin-orbit interaction, so called $L$-type MDMs appear. There are helical MS waves in a quasi-2D ferrite disk [9, 10]. The $L$ modes are microwave MDM polaritons [12]. These polariton states create near fields with specific topology – the magnetoelectric (ME) fields [13]. There are topologically distinctive virtual photons. As we well know, in a subwavelength region of regular microwave radiation the near-field structure is exhibited as a quasi-static electric or quasi-static magnetic field. These quasi-static electric and magnetic fields are mutually uncoupled [45]. The near-field structure of a MDM particle is essentially different. One can observe strong subwavelength localization of both the electric and magnetic fields [12, 13, 15, 40].

Topological properties of ME fields (non-zero helicity factor) arise from the presence of geometric phases on a border circle of a MDM ferrite disk. Due to geometrical phases, the in-plane power-flow vortices appear. It is evident that, in accordance with the thermodynamics laws, power cannot flow along a closed circle. So we have flat (Archimedes) spirals of the power-flow distributions. The incident-wave power flow goes from peripheral regions to the vortex center. The scattered-wave power flow goes from the vortex center to peripheral regions. There are or right-handed, or left-handed flat spirals. When, for example, the incident-wave power flow is a right-handed spiral, the scattered-wave power flow is a left-handed spiral. The *L*-mode solutions, being not the energy eigenstate solutions, are well observable in the HFSS numerical simulation. In a "dynamical" HFSS program, one can observe such geometric phases indirectly – by topological transformations of the wave fronts, subwavelength-scaled power flow vortices, and the helicity factors [11 – 13, 15, 40].

ME fields are the topological-defect solutions, which are distinct from the Maxwellian vacuum solutions. It can be proven to exist because the boundary conditions entail the existence of homotopically distinct solutions; the solutions to the differential equations are then topologically distinct. Experiments with a thin-film ferrite disk embedded in a microwave structure show that quantization of a microwave energy takes place due to the microwave-photon angular momentum. The discrete topological states of the microwave cavity fields are caused by discrete variation of energy of a ferrite disk appearing because of an external source of energy – a bias magnetic field [21]. The modes observed in a microwave cavity with an embedded MDM ferrite disk are quantum vacuum fluctuations [21].

## IV. EXPERIMENTAL AND NUMERICAL RESULTS

### A. Chiral edge magnetic currents

While the *L*-mode resonances are energetically unstable, there are topologically stable resonances of rotating fields [11 – 13]. These solutions are characterized by both the linear and circular magnetic currents. The two magnetic currents, being coupled at the MDM resonance, form helical-structure (or chiral) magnetic currents. It means that all excitations in a ferrite disk can propagate only clockwise or only counterclockwise. When, for a given direction of a bias magnetic field, we call the chiral-current process as a "forward" process in the formalism of fluctuation relations, in an opposite direction of a bias magnetic field we have a "backward" process. The backward process can be described as a forward process in the time-reversed twin system with the opposite chirality. Certainly, it should be assumed that the material characteristics and temperatures are the same in the forward and backward processes. In quantum-Hall structures, one can observe chiral excitations on a 2D surface of a 3D system. In such systems, electric charge can propagate in both directions along one of the coordinate axes but only in one direction along the second axis [14]. Similar situation, but with "magnetic charges", takes place in our case.

While positions of the MDM resonance peaks are exceptionally determined by the disk parameters, the amplitudes and forms of these peaks can be strongly dependent on the microwave-structure environment. The observed effect of multiresonant unidirectional tunneling (UDT) is due to such a dependence of the MDM spectrum on the microwave-structure properties. Based on the experimental and numerical studies, we show that breaking of symmetry in geometry of a microwave structure strongly influence on the UDT characteristics. The main physical aspect concerns the presence of chiral magnetic currents in a microwave structure. At the MDM resonances, these chiral magnetic currents result in unidirectional transfer of quantized angular momenta through subwavelength vacuum or isotropic-dielectric regions.

We start with a microstrip structure with an embedded thin-film ferrite disk. Such a microstrip structure is shown in Fig. 4. For experimental studies, we use a ferrite disk with the following parameters. The yttrium iron garnet (YIG) disk has a diameter of $D = 3$ mm and a thickness of $t = 0.05$ mm. The saturation magnetization of a ferrite is $4\pi M_s = 1880$ G. The linewidth of a ferrite is $\Delta H = 0.8$ Oe. The disk is normally magnetized by a bias magnetic field $H_0 = 4210$ Oe. An experimental microstrip structure is realized on a dielectric substrate (Taconic RF-35, $\varepsilon_r = 3.52$, thickness of 1.52 mm). Characteristic impedance of a microstrip line is 50 Ohm. The $S$-matrix parameters were measured by a network analyzer. With use of a current supply we established a quantity of a normal bias magnetic field $\vec{H}_0$, necessary to get the MDM spectrum in a required frequency range. For numerical studies, we use a ferrite disk with the same parameters as pointed above. The only difference is that, for better understanding the field structures, in numerical analyses we consider a ferrite disk with very small losses: the linewidth of a ferrite is $\Delta H = 0.1$ Oe. In the absence of MDM resonance peaks, the subwavelength coupling between two microstrip lines in these structures is extremely small. At the resonance peaks, one has strong subwavelength coupling. Taking into account in-plane geometry, the presence of a ground metallic plane in a microwave structure, and direction of rotation (at a given direction of a bias magnetic field) of a power-flow density in a ferrite disk, one finds that the ways electromagnetic waves propagating from port 1 to port 2 and, oppositely, from port 2 to port 1, are geometrically different.

Fig. 5 represents the experimental frequency characteristics of modules of the reflection coefficient (the $S_{11}$ scattering-matrix parameter) and the transmission coefficients (the $S_{21}$ and $S_{12}$ scattering-matrix parameters) for two opposite orientations of a normal bias magnetic field $\vec{H}_0$. Classification of the resonances shown in Fig. 5 is based on analytical studies in Ref. [8]. There are resonances corresponding to MDMs with radial and azimuth variations of the magnetostatic-potential wave functions in a ferrite disk. The azimuth-variation resonances appear because of the azimuth nonhomogeneity of a microstrip structure. In the spectra, these resonances are observed between the peaks of radial-variation resonances. In the mode designation, the first number characterizes a number of radial variations for the MDM spectral solution. The second number is a number of azimuth variations [8, 21].

In Fig. 5, one can see that the reflection-coefficient spectrum is characterized by the Lorentz-resonance peaks. Contrarily, the transmission-coefficient excitations manifest themselves as the Fano-resonance peaks at the stationary states of the MDM spectrum. It is worth noting that a character of the Fano interference is different for radial-variation and azimuth-variation MDMs [21, 40]. For the spectra in Fig. 5, we can state that while the reflection-coefficient spectrum is the same at two opposite orientations of a normal bias magnetic field $\vec{H}_0$, for the transmission-coefficient excitations there is strong sensitivity of the peak sizes on the direction of a bias magnetic field. At a very small transmission level for non-resonant frequencies (about $|S_{21}| = -25 dB$), one clearly observes resonance peaks of the UDT. For better illustration of this effect, in Fig. 6 we show the reflection and transmission spectra normalized to the background (when a bias magnetic field is zero) level of the microwave structure. An analysis of the transmission spectra shows that the observed UDT effect is due to the field chirality in an *entire* microwave structure. When, for a given direction of a bias magnetic field, one has a "forward" process, the "backward" process is exhibited as a "forward" process in the time-reversed twin system with the opposite chirality. As the sources of the fields with different chirality, there are edge magnetic currents $\left[\left(\vec{j}_s^{(m)}\right)_{\theta'}\right]^{(i)}$ and $\left[\left(\vec{j}_s^{(m)}\right)_{\theta'}\right]^{(s)}$ considered in the above model. The field chirality results in unidirectional transfer of angular momenta through a subwavelength vacuum and isotropic-dielectric regions. Simultaneous exchange between the "forward" and "backward" processes together with change of the time direction remains the system symmetry unbroken. This symmetry properties of the chiral states are well illustrated in Fig. 7. One can see complete coincidence between the spectra of the $S_{21}$ and $S_{12}$ scattering-matrix parameters for oppositely directed bias magnetic fields.

The shown above experimental results of the UDT effect are well verified numerically. In Fig. 8, one can see the numerically obtained transmission characteristics for the first MDM resonance at two opposite orientations of a normal bias magnetic field $\vec{H}_0$. It is necessary to note here that instead of a bias magnetic field used in experiments ($H_0 = 4210$ Oe), in the numerical studies we applied a higher-quantity bias magnetic field: $H_0 = 4434$ Oe. Use of such a higher quantity (giving us the same position of the resonance peak in the experiments and numerical studies and) is necessary because of non-homogeneity of an internal DC magnetic field in a real ferrite disk. A more detailed discussion on a role of non-homogeneity of an internal DC magnetic field in the MDM spectral characteristics can be found in Ref. [8].

To explain the UDT effect in this microstrip structure, we can use the following model. At the MDM resonance frequency, for a given direction of a bias magnetic field the incident and scattered waves near a ferrite disk have different types of flat spirals. While, for example, the incident wave is a right-handed spiral, the scattered wave should be a left-handed spiral. In an opposite direction of a bias magnetic field, we have a left-handed spiral for the incident wave and a right-handed spiral for the scattered wave. The UDT effect appears for the reason that in a nonsymmetrical microwave structure the interplay between the right-handed and left-handed spirals of the incident and scattered waves is different for different directions of a normal bias magnetic field.

Since the UDT appears due to distinguishable topology of the "forward" and "backward" excitations, this effect should be enhanced for a microwave structure with increased breaking of symmetry in geometry. Such a microstrip structure is shown in Fig 9. The symmetry breaking is increased by an inclined slot in one of conductive strip. Fig. 10 shows the experimental $S_{21}$ and $S_{12}$ scattering-matrix parameters of this structure for two opposite orientations of a normal bias magnetic field $\vec{H}_0$. There are the transmission spectra both non-normalized and normalized to the background level of the microwave structure. One can find that, compared to the previous results, the UDT effect is strongly enhanced in the structure with an inclined slot. These results are well verified numerically in Fig. 11.

### B. Chiral edge electric currents

At MDM resonances, chiral electric currents can be induced on a thin metal wire placed on a surface of a ferrite disk [52]. In a structure shown in Fig. 12, the electric field of a microstrip system causes a linear displacement of electric charges when interacting with a short piece of a wire. At the same time, the magnetic field of a MDM vortex causes a circulation of electric charges. Being combined, these two motions (which include translation and rotation) cause helical motion of electrons on a surface of a metal wire. Such helical waves result in observation of very peculiar field structures. Because of a chiral surface electric current, the electric and magnetic fields at the butt end of a wire electrode become not mutually oriented in vacuum at the angle of $90°$. While the mutually perpendicular components of the electric and magnetic fields give the power-flow-density vortex, the mutually parallel components result in appearance of nonzero helicity density $F = \dfrac{1}{16\pi} \operatorname{Im} \left\{ \vec{E} \cdot \left( \vec{\nabla} \times \vec{E} \right)^* \right\}$ [52]. Fig. 13 shows numerically obtained distributions of the fields and currents on a wire electrode and also the field structures near a butt end of a wire electrode.

The field chirality in a MDM microwave structure with a wire electrode results in unidirectional transfer of angular momenta through a subwavelength vacuum region. Our experimental results give evidence for such a tunneling effect. To enhance experimental observation of the UDT effect due to chiral electric currents, we use structures with breaking of symmetry in geometry. There are the right- or left-handed metallic helices. Fig. 14 shows two-port microwave structures with a MDM ferrite disk and a wire electrode (port 1) and with the right- or left-handed metallic helices (port 2). In these structures we used a wire electrode with diameter of 100 um. Metallic helices are made with the same wire. Diameters of the helices are 2mm. The pitch is equal to 0.4mm. Every helix has ten turns. A wire concentrator is placed near a metallic helix without an electric contact with it.

Fig. 15 shows experimental results of the $S_{21}$ scattering-matrix parameter for the right- and left-handed metallic helices and two opposite directions of a bias magnetic field. The transmission spectra is normalized to the background (when a bias magnetic field is zero) level of the microwave structure. This background level is about $|S_{21}| = -30 dB$. It is evident that there is a specific chiral symmetry. Simultaneous change of the helix handedness and direction of bias magnetic field remains the system symmetry unbroken. Numerical results the transmission spectra shown in Fig. 16 are in good correspondence with experimental results.

The observed effect of unidirectional tunneling can be well explained by an analysis of the power-flow-density distributions in a vacuum region near a wire concentrator and a metallic helix. Such distributions are shown in Fig. 17 for two resonance peaks corresponding to the $1^{st}$ MDM – the peaks A and B in the $S_{21}$ frequency characteristics in Fig. 16. One can clearly see that the power transmission in a two-port microwave structure is maximal when a direction of the power-flow vortex at a butt end of a wire electrode corresponds to the handedness of a metallic helix. There is an evidence for the presence of the orbital-angular-momentum twisting excitations in a subwavelength region of microwave radiation at the MDM resonances. Due to chiral properties of the fields near a wire electrode, one has unidirectional transfer of quantized angular momenta through a subwavelength vacuum region.

In our previous studies [13, 15, 40, 52], it was shown that topology of the near fields originated from a MDM ferrite disk – the ME fields – is characterized both by the power-flow vortices and the helicity parameters. While the power transmission in a two-port microwave structure is strongly related to the power-flow vortices of the twisting excitations, the helicity characteristics of these excitations (being local characteristics of the field geometry) are very slightly correlated with the power transmission effect. Fig. 18 shows distributions of the normalized helicity factor for the two resonance peaks corresponding to the $1^{st}$ MDM – the peaks A and B in the $S_{21}$ frequency characteristics in Fig. 16. The normalized helicity parameter is defined as [13, 15, 40, 52]

$$\cos\alpha = \frac{\text{Im}\left\{\vec{E}\cdot\left(\vec{\nabla}\times\vec{E}\right)^*\right\}}{\left|\vec{E}\right|\left|\nabla\times\vec{E}\right|}. \tag{27}$$

In Fig. 18, one can see the helicity parameter distributions are mainly related to a direction of a bias magnetic field and periodicity of the metal-helix turns. There is very small relation between the helicity parameter distribution and the helix handedness.

## V. DISCUSSION AND CONCLUSION

The effects of quantum coherence involving macroscopic degrees of freedom and occurring in systems far larger than individual atoms are one of the topical fields in modern physics [53]. Recently, much progress has been made in demonstrating the macroscopic quantum behavior of superconductor systems, where particles form highly correlated electron system. The concept of coherent mixtures of electrons and holes, underlying the BdG-hamiltonian quasiclassical approximation, well describes the topological superconductors.

Macroscopic quantum coherence can also be observed in some ferrimagnetic structures. MDM oscillations in a quasi-2D ferrite disk are macroscopically (mesoscopically) quantized states. Long range dipole-dipole correlation in position of electron spins in a ferrimagnetic sample with saturation magnetization can be treated in terms of collective excitations of the system as a whole. If the sample is sufficiently small so that the dephasing length $L_{ph}$ of the magnetic dipole-dipole interaction exceeds the sample size, this interaction is non-local on the scale of $L_{ph}$. This is a feature of a mesoscopic ferrite sample, i.e., a sample with linear dimensions smaller than $L_{ph}$ but still much larger than the exchange-interaction scales. In a case of a quasi-2D ferrite disk, the quantized forms of these collective matter oscillations – magnetostatic magnons – were found to be quasiparticles with both wave-like and particle-like behavior, as expected for quantum excitations. The magnon motion in this system is quantized in the direction perpendicular to the plane of a ferrite disk. The MDM oscillations in a quasi-2D ferrite disk, analyzed as spectral solutions for the magnetostatic-potential wave function $\psi(\vec{r}, t)$, has evident quantum-like attributes. The discrete energy eigenstates of the MDM oscillations in a quasi-2D ferrite disk are well observed in the first microwave experiments [35, 36]. Experimental studies of interaction of the MDM ferrite particles with its microwave environment give evidence for multiresonance Fano-type interference characteristic for quantum structures [21, 37]. It was also shown that MDMs in a ferrite disk are topologically distinctive eigenstates [13, 15]. MDM oscillations in a quasi-2D ferrite disk can conserve energy and angular momentum. Because of these properties, MDMs strongly confine microwave radiation energy in subwavelength scales. In a vacuum subwavelength region abutting to a MDM ferrite disk, one can observe the quantized-state power-flow vortices. In such a region, a coupling between the time-varying electric and magnetic fields is different from such a coupling in regular electromagnetic fields. These specific near fields – the ME fields – give evidence for spontaneous symmetry breakings at the resonant states of MDM oscillations. This symmetry breaking is characterized by the helicity parameter[13, 15, 52].

The ME-field singularity is strongly related to edge chiral currents. At the MDM resonances one has chiral magnetic currents on a lateral surface of a ferrite disk. Also, in a MDM ferrite disk with a wire electrode, one can observe chiral electric currents on a surface of a metal wire. Because of the presence edge chiral currents, interaction of MDMs with external microwave radiation is characterized by unique topological properties. In this paper, we showed that due to a topological structure of ME fields one obtains unidirectional multiresonant tunneling for electromagnetic waves propagating in microwave structures with embedded quasi-2D ferrite disks. The resonances manifest themselves as peaks in the scattering-matrix parameters at the stationary states of MDM oscillations in a ferrite disk. The effect of unidirectional tunneling is exhibited as the Fano-resonance interferences, which are different for oppositely directed bias magnetic fields. The observed multiresonant tunneling is due to generation of quantized vortices. The resulting quantum vacuum torque is strong enough to provide a contactless transfer of angular momentum to the helical-form metallic sample.

The MDM twisting excitations in a subwavelength region of microwave radiation in two-port structures can be well described by the formula for the scattering-matrix parameters:

$$S_{12}^{H_0\uparrow} = S_{21}^{H_0\downarrow} \quad \& \quad S_{21}^{H_0\uparrow} = S_{12}^{H_0\downarrow} \ . \tag{28}$$

The observed chiral fields are near fields which are fundamentally different from the fields radiated by usual microwave antennas with ferrite inclusions. In the last case, the radiation characteristics do not depend on a direction of a bias field. So the $S$-matrix parameters of the two-port (transmitter-receiver) structure do not depend on a direction of a bias field. In usual microwave antennas with ferrite inclusions we have symmetry of the $S$ matrix: $S_{ij} = S_{ji}$, $i \neq j$.

## Refrerences

**Figure captions**

Fig. 1. Surface magnetic charges and edge chiral magnetic currents (a view on the upper plane of a ferrite disk). At the time phase variation from $\omega t = 0$ to $\omega t = \pi$, an edge magnetic current acquires the phase of $\theta' = \pi$ while the $G$-mode regular-coordinate angle is $\theta = 2\pi$. Because of a magnetic-dipole fluctuation on a lateral surface of a ferrite disk the domain of the $G$-mode azimuthal angle $\theta$, in a laboratory frame, is no more $[0, 2\pi]$ but $[0, 4\pi]$. The MS-potential distribution for the $G$-mode eigenfunction $\tilde{\eta}$ is schematically shown as color regions inside a ferrite disk. In the figure, there is a correspondence between colors used for surface magnetic currents and colors used for topological magnetic charges.

Fig. 2. A two-layer-ring model for edge chiral magnetic currents for different time phases. When a magnetic current of the upper (lower) layer is arriving to terminal $P$ (where a topological magnetic charge is localized), it must continue its propagation at the lower (upper) layer and this is only one choice. The similar situation takes place at terminal $Q$. Regions of terminals $P$ and $Q$ are the regions of singularity (the regions of topological magnetic charges). Topological magnetic charges are distributed on a lateral surface of a ferrite disk. In the figure, there is a correspondence between colors used for surface magnetic currents and colors used for topological magnetic charges.

Fig. 3. Edge magnetic currents $\left[\left(\vec{j}_s^{(m)}\right)_{\theta'}\right]^{(i)}$ and $\left[\left(\vec{j}_s^{(m)}\right)_{\theta'}\right]^{(s)}$ on a lateral surface of a ferrite disk and their correlation with the $G$-mode MS-potential wave functions $\tilde{\eta}$ of a certain MDM. The spin-orbit interaction is illustrated by highlighted parts of in the graphs of the currents $\left[\left(\vec{j}_s^{(m)}\right)_{\theta'}\right]^{(i)}$ and $\left[\left(\vec{j}_s^{(m)}\right)_{\theta'}\right]^{(s)}$ and shaded areas on the graph of the wave functions $\tilde{\eta}$. Every of the magnetic currents, $\left[\left(\vec{j}_s^{(m)}\right)_{\theta'}\right]^{(i)}$ and $\left[\left(\vec{j}_s^{(m)}\right)_{\theta'}\right]^{(s)}$, is composed by two topologically distinctive current components shown in Figs. 1 and 2.

Fig. 4. A microstrip structure with an embedded thin-film ferrite disk.

Fig. 5. Experimental evidence for unidirectional multiresonant tunneling. Frequency characteristics of modules of the scattering-matrix parameters for two opposite orientations of a normal bias magnetic field $\vec{H}_0$. (*a*) The reflection coefficient; (*b*), (*c*) the transmission coefficients. The resonances are classified based on analytical studies in Ref. [8]. The first number characterizes a number of radial variations for the MDM spectral solution. The second number is a number of azimuth variations.

Fig. 6. The reflection and transmission spectra the same as in Fig. 5, but normalized to the background (when a bias magnetic field is zero) level of the microwave structure. (*a*) The reflection coefficient; (*b*), (*c*) the transmission coefficients.

Fig. 7. Symmetry properties of the chiral states in a microwave structure with a MDM ferrite disk. There is complete coincidence between the spectra of the $S_{21}$ and $S_{12}$ scattering-matrix parameters for oppositely directed bias magnetic fields. Microwave radiation in two-port structure can be described by the formula for the scattering-matrix parameters: $S_{12}^{H_0\uparrow} = S_{21}^{H_0\downarrow}$ & $S_{21}^{H_0\uparrow} = S_{12}^{H_0\downarrow}$ .

Fig. 8. The numerically obtained transmission characteristics for the first MDM resonance at two opposite orientations of a normal bias magnetic field $\vec{H}_0$. (*a*) The $S_{21}$ scattering-matrix parameter; (*b*) the $S_{12}$ scattering-matrix parameter.

Fig 9. A microwave structure with increased breaking of symmetry in geometry. The symmetry breaking is increased by an inclined slot in one of conductive strip.

Fig. 10. The experimental $S_{21}$ and $S_{12}$ scattering-matrix parameters of the structure shown in Fig. 9 for two opposite orientations of a normal bias magnetic field $\vec{H}_0$. (*a*), (*b*) Non-normalized transmission spectra; (*c*), (*d*) Transmission spectra normalized to the background level of the microwave structure.

Fig. 11. The numerically obtained transmission characteristics for a microwave structure with increased breaking of symmetry in geometry. (*a*) The $S_{21}$ scattering-matrix parameter; (*b*) the $S_{12}$ scattering-matrix parameter.

Fig. 12.   (a) A microstrip structure with a MDM ferrite disk and a wire electrode. (b) A magnified picture of a ferrite disk and a wire.

Fig. 13. Distributions of the fields and currents on a wire electrode and the field structures near a butt end of a wire electrode. (*a*) Electric field on a wire electrode; (*b*) surface electric chiral current; (*c*) and (*d*) electric and magnetic fields near a butt end of a wire electrode; (*e*) and (*f*) power-flow density and helicity density near a butt end of a wire electrode.

Fig. 14. (a) Two-port microwave structures with a MDM ferrite disk and a wire electrode (port 1) and with the right- or left-handed metallic helices (port 2). (b) A magnified picture. A wire concentrator is placed near a metallic helix without an electric contact with it.

Fig. 15. Experimental evidence for unidirectional multiresonance tunneling due to chiral edge electric currents. Frequency characteristics of the $S_{21}$ scattering-matrix parameter for two opposite orientations of a normal bias magnetic field $\vec{H}_0$. (*a*) The right- handed metallic helix; (*b*) the left-handed metallic helix. The transmission spectra is normalized to the background (when a bias magnetic field is zero) level of the microwave structure. The background level is about $|S_{21}| = -30 dB$. The system has chiral symmetry: simultaneous change of the helix handedness and direction of bias magnetic field remains the system symmetry unbroken.

Fig. 16. Numerical results of the $S_{21}$ scattering-matrix parameter for two opposite orientations of a normal bias magnetic field $\vec{H}_0$. (*a*) The right-handed metallic helix; (*b*) the left-handed metallic helix.

Fig. 17. The power-flow-density distributions in a vacuum region near a wire concentrator and a metallic helix shown for two resonance peaks corresponding to the 1$^{st}$ MDM – the peaks A and B in the $S_{21}$ frequency characteristics in Fig. 16. (*a*) and (*b*) the right- and left-handed metallic helices at a normal bias magnetic field $\vec{H}_0$ directed upwards; (*c*) and (*d*) the right- and left-handed metallic helices at a normal bias magnetic field $\vec{H}_0$ directed downwards. The power transmission in a two-port microwave structure is maximal when a direction of the power-flow vortex at a butt end of a wire electrode corresponds to the handedness of a metallic helix. There is an evidence for the presence of the orbital-angular-momentum twisting excitations in a subwavelength region of microwave radiation at the MDM resonances.

Fig. 18. Distributions of the normalized helicity factor for the two resonance peaks corresponding to the 1$^{st}$ MDM – the peaks A and B in the $S_{21}$ frequency characteristics in Fig. 16. (*a*) and (*b*) the right- and left-handed metallic helices at a normal bias magnetic field $\vec{H}_0$ directed upwards; (*c*) and (*d*) the right- and left-handed metallic helices at a normal bias magnetic field $\vec{H}_0$ directed downwards.

----------------------------------------------------------------------------------------------------------------------

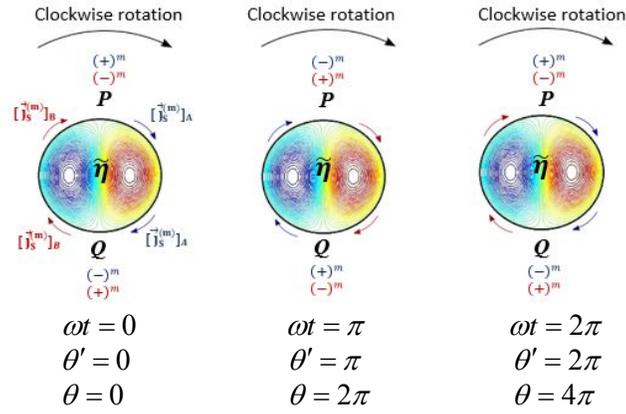

Fig. 1. Surface magnetic charges and edge chiral magnetic currents (a view on the upper plane of a ferrite disk). At the time phase variation from $\omega t = 0$ to $\omega t = \pi$, an edge magnetic current acquires the phase of $\theta' = \pi$ while the $G$-mode regular-coordinate angle is $\theta = 2\pi$. Because of a magnetic-dipole fluctuation on a lateral surface of a ferrite disk the domain of the $G$-mode azimuthal angle $\theta$, in a laboratory frame, is no more $[0, 2\pi]$ but $[0, 4\pi]$. The MS-potential distribution for the $G$-mode eigenfunction $\tilde{\eta}$ is schematically shown as color regions inside a ferrite disk. In the figure, there is a correspondence between colors used for surface magnetic currents and colors used for topological magnetic charges.

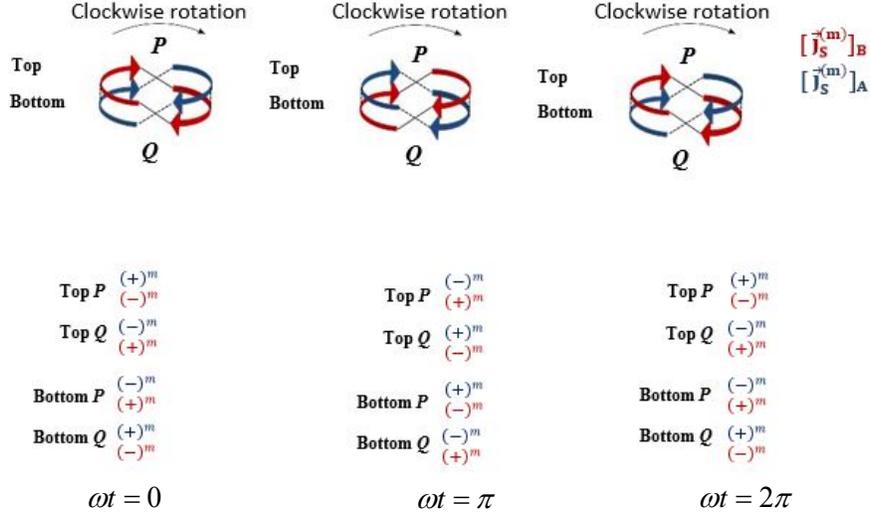

Fig. 2. A two-layer-ring model for edge chiral magnetic currents for different time phases. When a magnetic current of the upper (lower) layer is arriving to terminal $P$ (where a topological magnetic charge is localized), it must continue its propagation at the lower (upper) layer and this is only one choice. The similar situation takes place at terminal $Q$. Regions of terminals $P$ and $Q$ are the regions of singularity (the regions of topological magnetic charges). Topological magnetic charges are distributed on a lateral surface of a ferrite disk. In the figure, there is a correspondence between colors used for surface magnetic currents and colors used for topological magnetic charges.

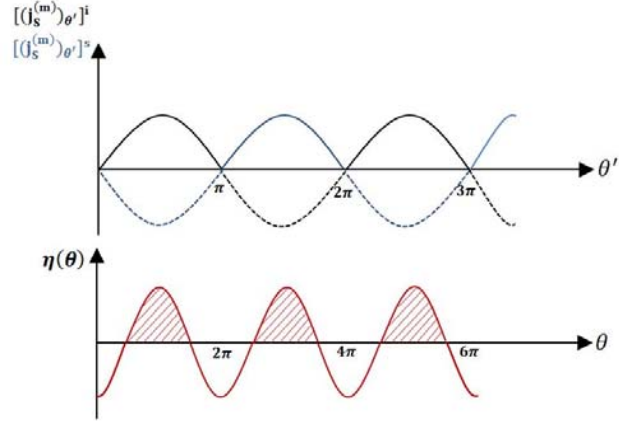

Fig. 3. Edge magnetic currents $\left[\left(\vec{j}_s^{(m)}\right)_{\theta'}\right]^{(i)}$ and $\left[\left(\vec{j}_s^{(m)}\right)_{\theta'}\right]^{(s)}$ on a lateral surface of a ferrite disk and their correlation with the $G$-mode MS-potential wave functions $\tilde{\eta}$ of a certain MDM. The spin-orbit interaction is illustrated by highlighted parts of in the graphs of the currents $\left[\left(\vec{j}_s^{(m)}\right)_{\theta'}\right]^{(i)}$ and $\left[\left(\vec{j}_s^{(m)}\right)_{\theta'}\right]^{(s)}$ and shaded areas on the graph of the wave functions $\tilde{\eta}$. Every of the magnetic currents, $\left[\left(\vec{j}_s^{(m)}\right)_{\theta'}\right]^{(i)}$ and $\left[\left(\vec{j}_s^{(m)}\right)_{\theta'}\right]^{(s)}$, is composed by two topologically distinctive current components shown in Figs. 1 and 2.

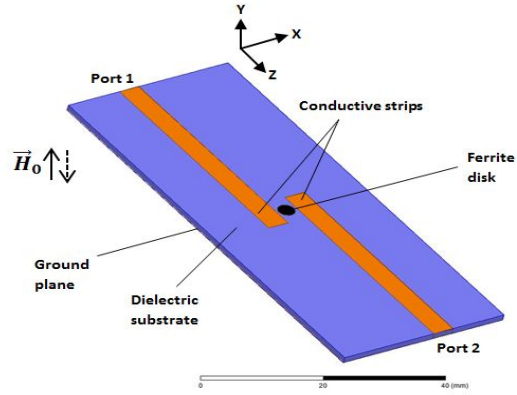

Fig. 4. A microstrip structure with an embedded thin-film ferrite disk.

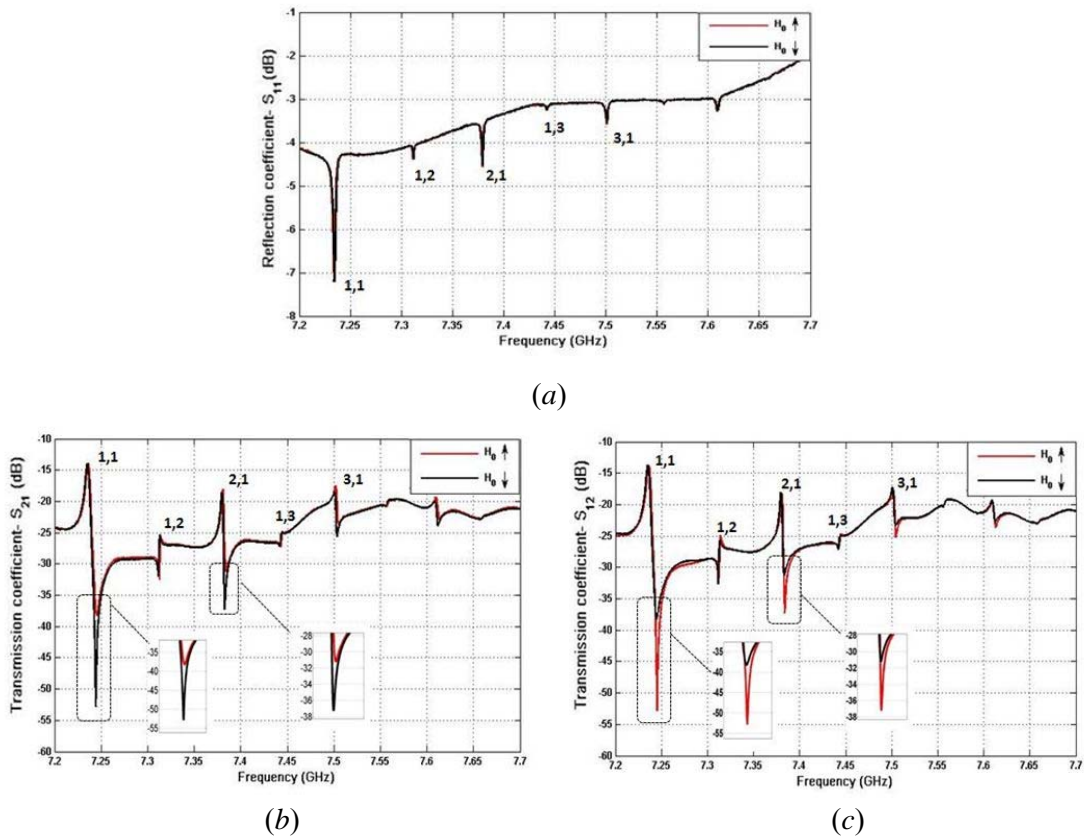

Fig. 5. Experimental evidence for unidirectional multiresonance tunneling. Frequency characteristics of modules of the scattering-matrix parameters for two opposite orientations of a normal bias magnetic field $\vec{H}_0$. (*a*) The reflection coefficient; (*b*), (*c*) the transmission coefficients. The resonances are classified based on analytical studies in Ref. [8]. The first number characterizes a number of radial variations for the MDM spectral solution. The second number is a number of azimuth variations.

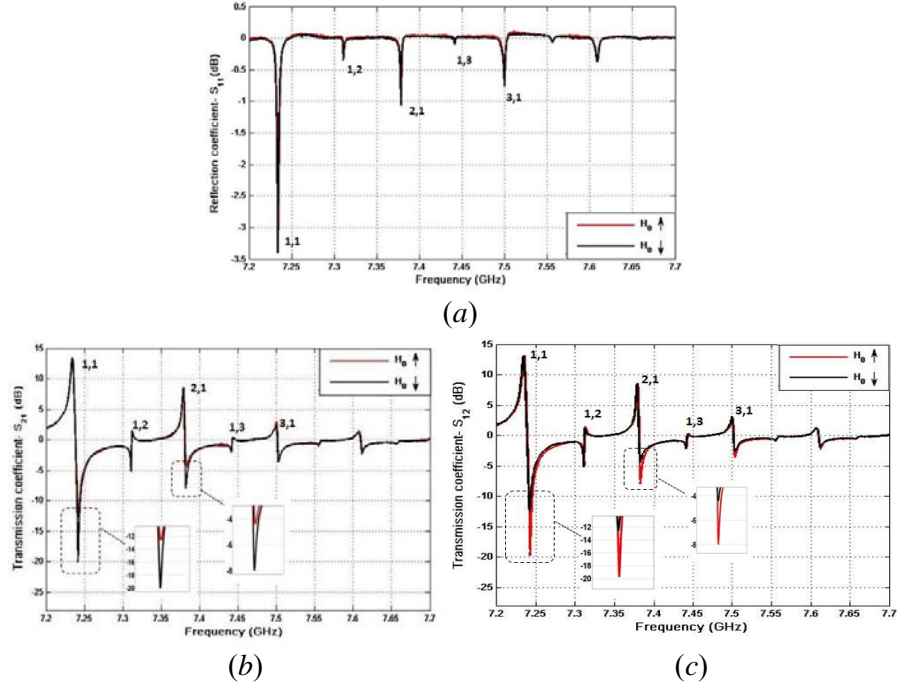

(a)

(b)                                    (c)

Fig. 6. The reflection and transmission spectra the same as in Fig. 5, but normalized to the background (when a bias magnetic field is zero) level of the microwave structure. (a) The reflection coefficient; (b), (c) the transmission coefficients.

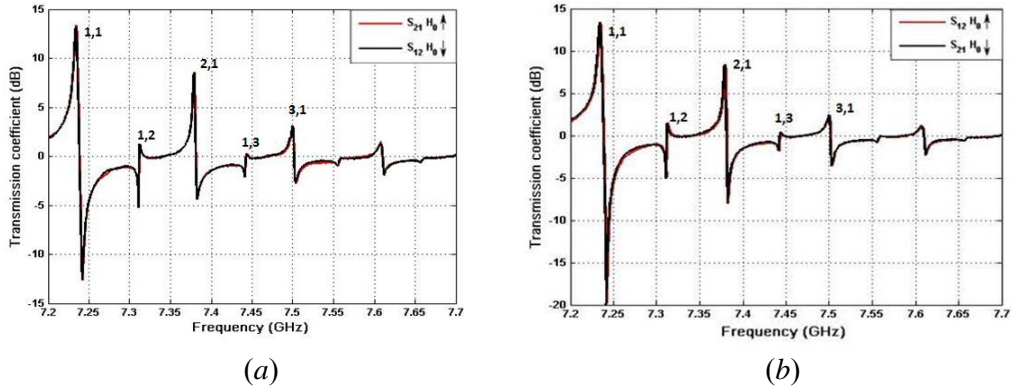

(a)                                    (b)

Fig. 7. Symmetry properties of the chiral states in a microwave structure with a MDM ferrite disk. There is complete coincidence between the spectra of the $S_{21}$ and $S_{12}$ scattering-matrix parameters for oppositely directed bias magnetic fields. Microwave radiation in two-port structure can be described by the formula for the scattering-matrix parameters: $S_{12}^{H_0\uparrow} = S_{21}^{H_0\downarrow}$ & $S_{21}^{H_0\uparrow} = S_{12}^{H_0\downarrow}$ .

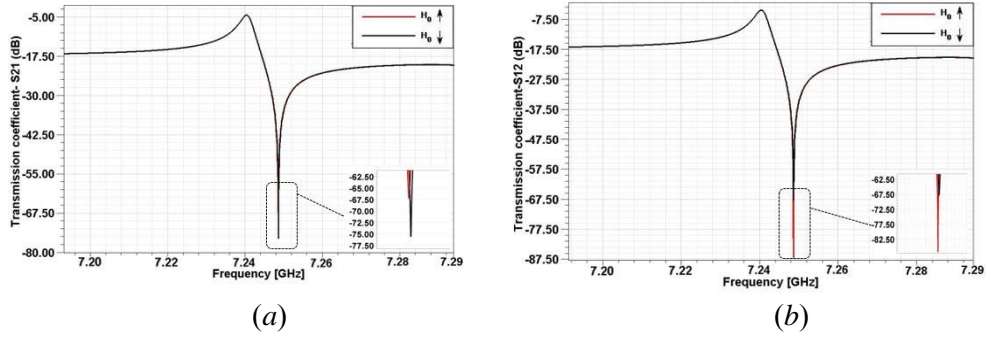

(a)                  (b)

Fig. 8. The numerically obtained transmission characteristics for the first MDM resonance at two opposite orientations of a normal bias magnetic field $\vec{H}_0$. (a) The $S_{21}$ scattering-matrix parameter; (b) the $S_{12}$ scattering-matrix parameter.

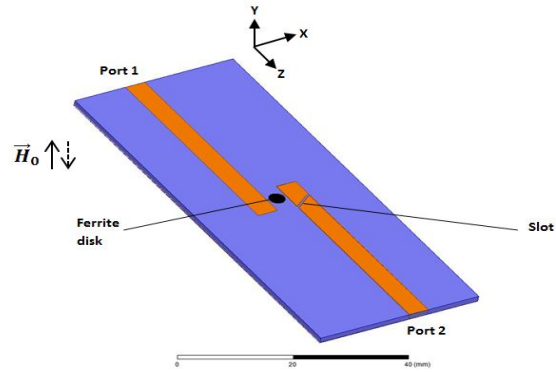

Fig 9. A microwave structure with increased breaking of symmetry in geometry. The symmetry breaking is increased by an inclined slot in one of conductive strip.

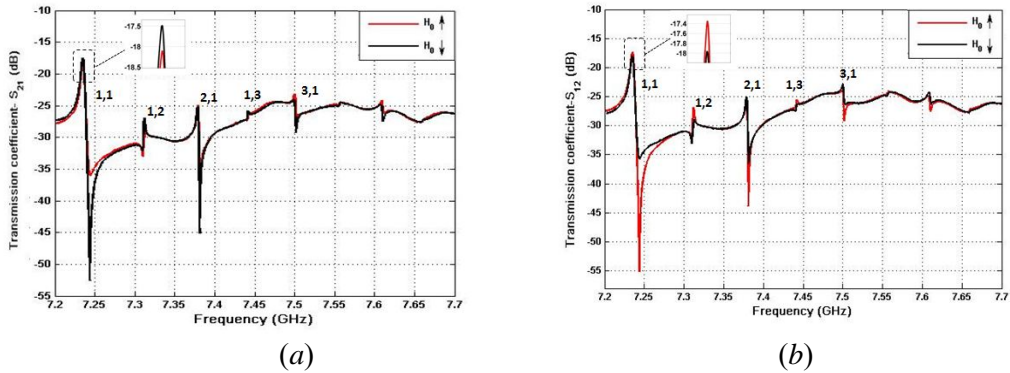

(a)                  (b)

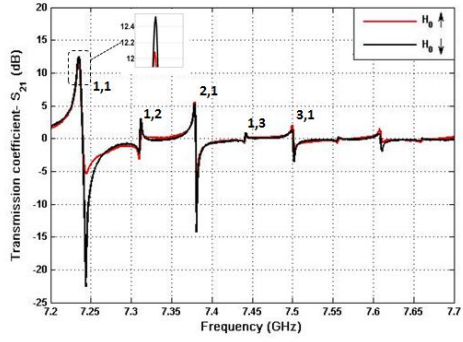

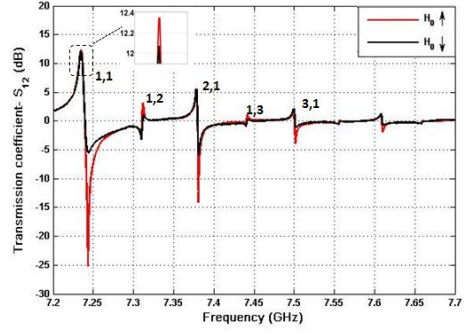

<div align="center">(<i>c</i>)          (<i>d</i>)</div>

Fig. 10. The experimental $S_{21}$ and $S_{12}$ scattering-matrix parameters of the structure shown in Fig. 9 for two opposite orientations of a normal bias magnetic field $\vec{H}_0$. (*a*), (*b*) Non-normalized transmission spectra; (*c*), (*d*) Transmission spectra normalized to the background level of the microwave structure.

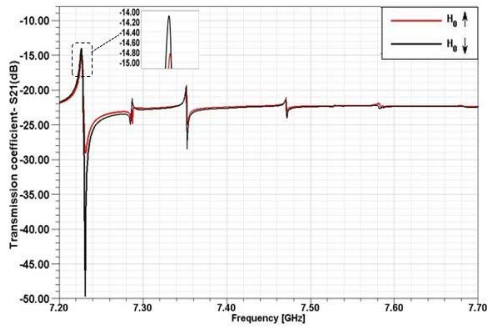

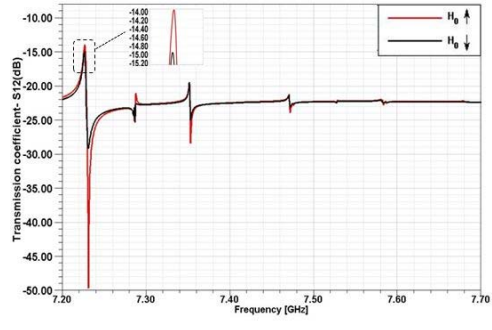

<div align="center">(<i>a</i>)          (<i>b</i>)</div>

Fig. 11. The numerically obtained transmission characteristics for a microwave structure with increased breaking of symmetry in geometry. (*a*) The $S_{21}$ scattering-matrix parameter; (*b*) the $S_{12}$ scattering-matrix parameter.

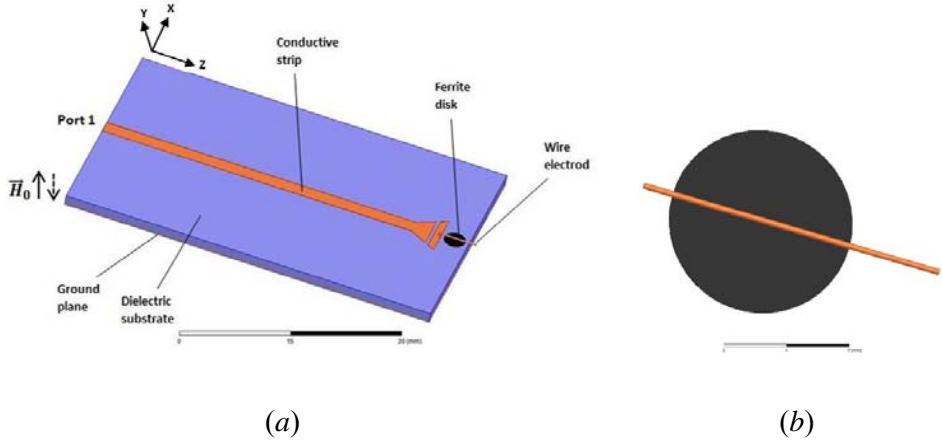

<div align="center">(<i>a</i>)          (<i>b</i>)</div>

Fig. 12. (*a*) A microstrip structure with a MDM ferrite disk and a wire electrode. (*b*) A magnified picture of a ferrite disk and a wire.

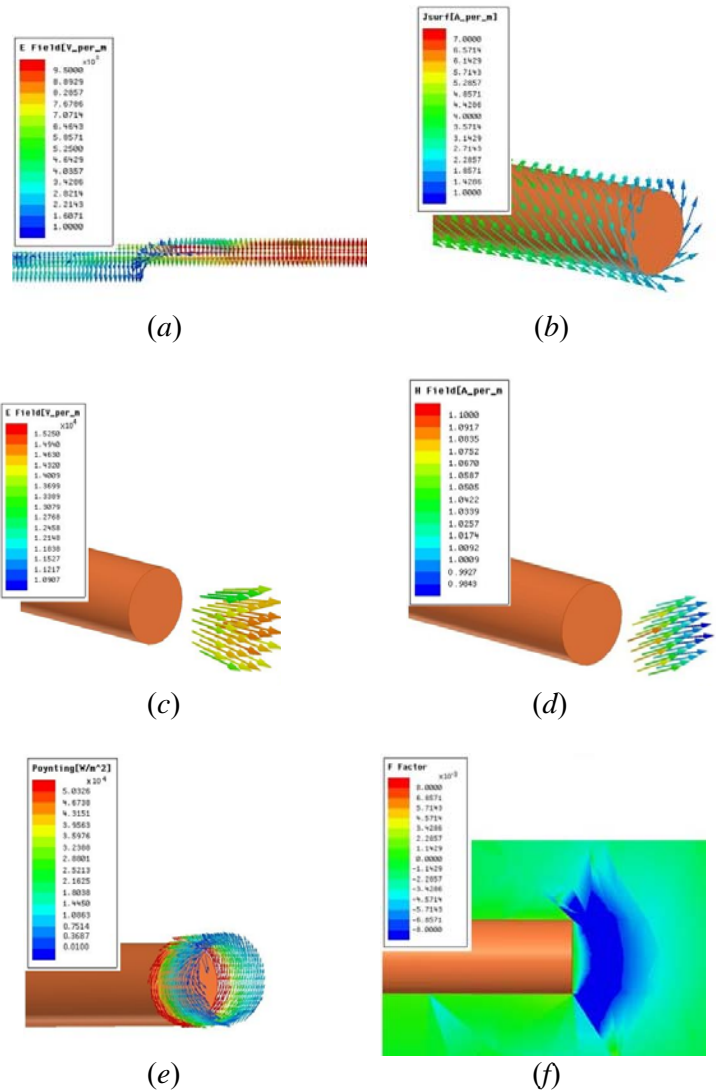

Fig. 13. Distributions of the fields and currents on a wire electrode and the field structures near a butt end of a wire electrode. (*a*) Electric field on a wire electrode; (*b*) surface electric chiral current; (*c*) and (*d*) electric and magnetic fields near a butt end of a wire electrode; (*e*) and (*f*) power-flow density and helicity density near a butt end of a wire electrode.

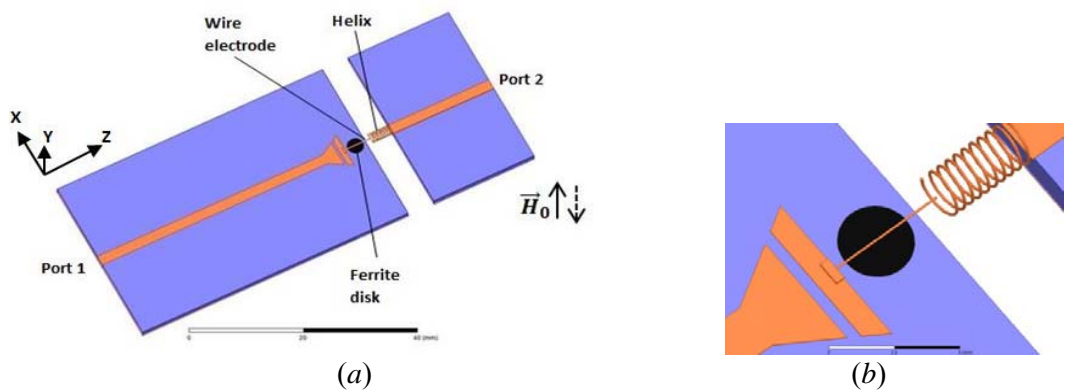

Fig. 14. (a) Two-port microwave structures with a MDM ferrite disk and a wire electrode (port 1) and with the right- or left-handed metallic helices (port 2). (b) A magnified picture. A wire concentrator is placed near a metallic helix without an electric contact with it.

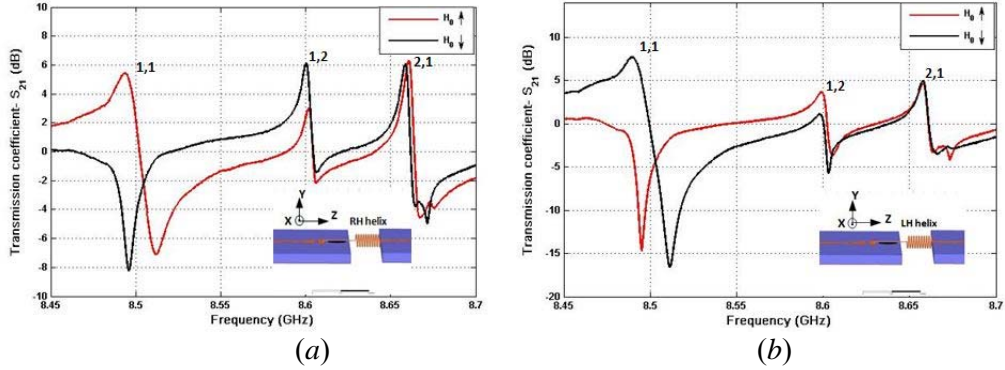

(a)                                         (b)

Fig. 15. Experimental evidence for unidirectional multiresonance tunneling due to chiral edge electric currents. Frequency characteristics of the $S_{21}$ scattering-matrix parameter for two opposite orientations of a normal bias magnetic field $\vec{H}_0$. (a) The right-handed metallic helix; (b) the left-handed metallic helix. The transmission spectra is normalized to the background (when a bias magnetic field is zero) level of the microwave structure. The background level is about $\left|S_{21}\right| = -30 dB$. The system has chiral symmetry: simultaneous change of the helix handedness and direction of bias magnetic field remains the system symmetry unbroken.

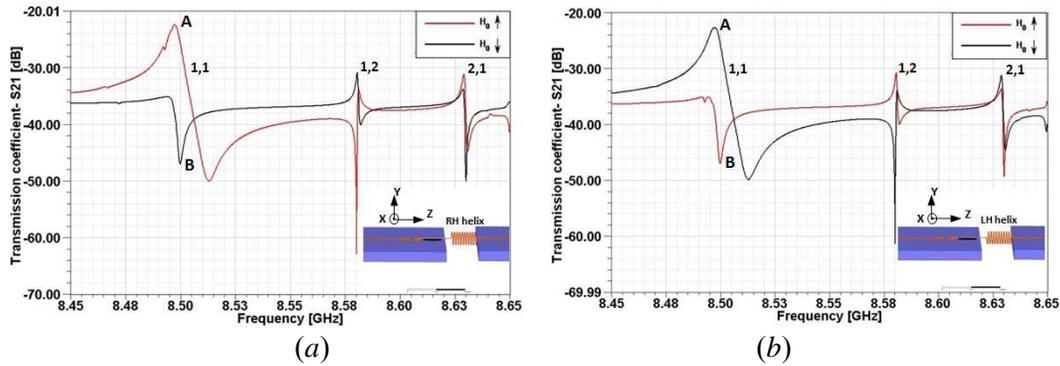

(a)                                         (b)

Fig. 16. Numerical results of the $S_{21}$ scattering-matrix parameter for two opposite orientations of a normal bias magnetic field $\vec{H}_0$. (a) The right-handed metallic helix; (b) the left-handed metallic helix.

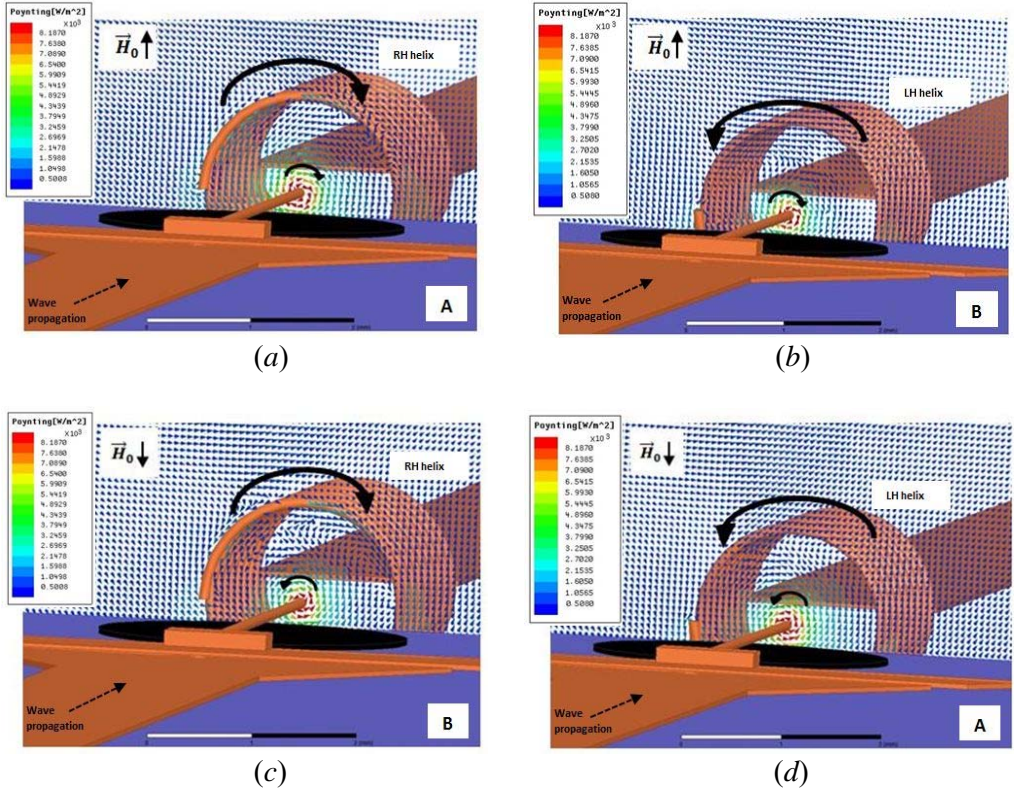

(a)                                          (b)

(c)                                          (d)

Fig. 17. The power-flow-density distributions in a vacuum region near a wire concentrator and a metallic helix shown for two resonance peaks corresponding to the 1$^{st}$ MDM – the peaks A and B in the $S_{21}$ frequency characteristics in Fig. 16. (a) and (b) the right- and left-handed metallic helices at a normal bias magnetic field $\vec{H}_0$ directed upwards; (c) and (d) the right- and left-handed metallic helices at a normal bias magnetic field $\vec{H}_0$ directed downwards. The power transmission in a two-port microwave structure is maximal when a direction of the power-flow vortex at a butt end of a wire electrode corresponds to the handedness of a metallic helix. There is an evidence for the presence of the orbital-angular-momentum twisting excitations in a subwavelength region of microwave radiation at the MDM resonances.

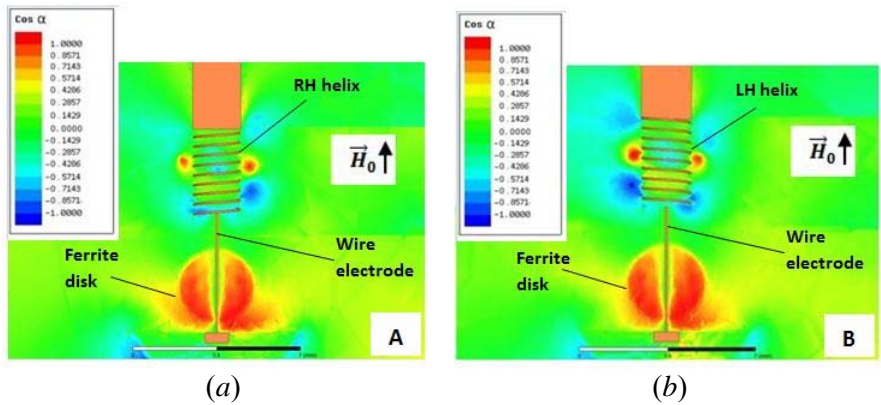

(a)                                          (b)

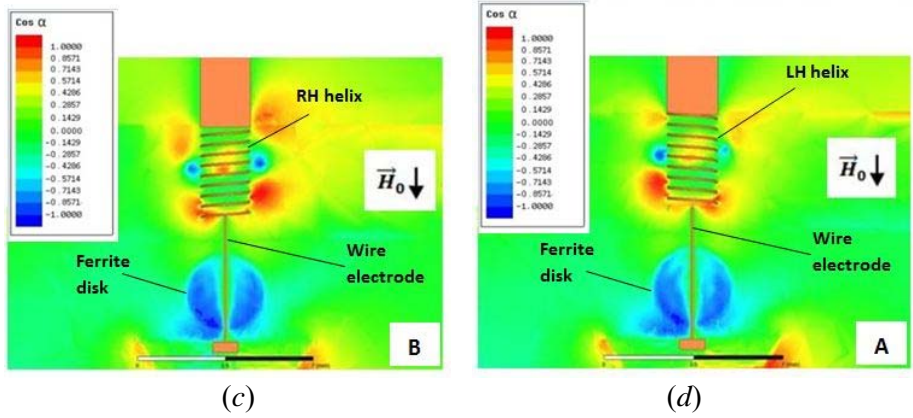

Fig. 18. Distributions of the normalized helicity factor for the two resonance peaks corresponding to the 1$^{st}$ MDM – the peaks A and B in the $S_{21}$ frequency characteristics in Fig. 16. (*a*) and (*b*) the right- and left-handed metallic helices at a normal bias magnetic field $\vec{H}_0$ directed upwards; (*c*) and (*d*) the right- and left-handed metallic helices at a normal bias magnetic field $\vec{H}_0$ directed downwards.